# Giant plateau-like topological Hall effect controlled by tailoring the magnetic exchange stiffness in a kagome magnet

Wei Xia,[1, 11, †] Aile Wang,[3, †] Jian Yuan,[1, †] Jiawei Luo,[1, 11] Yurui Wei,[1] Haonan Wang,[4] Wenjie Meng,[5] Yubin Hou,[5] Hong Du,[6] Xiangqi Liu,[1] Jiangteng Guo,[7] Yixuan Luo,[1] Ke Qu,[4] Min Chen,[1] Jinlong Jiao,[9] Xia Wang,[1, 8] Xuerong Liu,[1, 10] Wenbo Wang,[1, 11] Yulin Chen,[11, 13] Jianpeng Liu,[1, 11] Xuewen Fu,[7] Ruidan Zhong,[6, *] Qingyou Lu,[3, 5, 14, *] Shihao Zhang,[2, *] Zhenzhong Yang,[4, *] and Yanfeng Guo[1, 11, *]

[1]School of Physical Science and Technology, ShanghaiTech University, Shanghai 201210, China

[2]School of Physics and Electronics, Hunan University, Changsha 410082, China

[3]Hefei National Research Center for Physics Sciences at the Microscale, University of Science and Technology of China; Hefei, China

[4]School of Physics and Electronic Science, East China Normal University, Shanghai 200241 China

[5]Anhui Key Laboratory of Low-Energy Quantum Materials and Devices, High Magnetic Field Laboratory, HFIPS, Chinese Academy of Sciences, Hefei, China

[6]Tsung-Dao Lee Institute, Shanghai Jiao Tong University, Shanghai 201210, China

[7]Ultrafast Electron Microscopy Laboratory, The MOE Key Laboratory of Weak-Light Nonlinear Photonics School of Physics, Nankai University Tianjin 300071, China

[8]Analytical Instrumentation Center, School of Physical Science and Technology, ShanghaiTech University, Shanghai 201210, China

[9]Key Laboratory of Artificial Structures and Quantum Control (Ministry of Education), School of Physics and Astronomy, Shanghai Jiao Tong University, Shanghai 200240, China

[10]Center for Transformative Science, ShanghaiTech University, Shanghai 201210, China

[11]ShanghaiTech Laboratory for Topological Physics, ShanghaiTech University, Shanghai 201210, China

[12]Clarendon Laboratory, Department of Physics, University of Oxford, Oxford OX1 3PU, United Kingdom

[13]Anhui Laboratory of Advanced Photon Science and Technology, University of Science and Technology of China, University of Science and Technology of China; Hefei, China

[†]These authors contributed equally to this work:

Wei Xia, Aile Wang, and Jian Yuan.

[*]Correspondence:

*rzhong@sjtu.edu.cn

*qxl@ustc.edu.c

*zhangshh@hnu.edu.cn

*zzyang@phy.ecnu.edu.cn

*guoyf@shanghaitech.edu.cn

**The ferrimagnet $TbMn_6Sn_6$ has attracted vast attention, because its pristine Mn kagome lattice with strong spin-orbit coupling and out-of-plane Tb-Mn exchange supports quantum-limit Chern topological magnetism which can be described by the simple spinless Haldane model. We unveil herein that engineering the kagome lattice through partial substitution of Mn with nonmagnetic Cr induces a striking structural reorganization—Cr preferentially concentrates within a single Mn layer per unit cell, reducing the crystal symmetry from the $D_{6h}$ point group to the $C_2$. This tailored structure configuration gives rise to a plateau-like topological Hall effect (THE), achieving a record-breaking resistivity of 19.1 μΩ·cm among bulk systems. Complementary magnetic force microscopy measurements unveil a magnetic domain transition near 1 T at 180 K, aligning with the field-dependent phase diagram of the THE. Our direct visualization of the magnetic domain structure underscores the critical role of broken kagome lattice symmetry in** **generating distinct exchange stiffness between the two Mn layers. These findings establish a new paradigm for exploring exotic states in kagome topological magnets and provide a proof-of-principle strategy for unraveling the interplay between magnetism and emergent topological properties in kagome systems.**

## Introduction

The Haldane model predicts a Chern insulator phase that can be realized in a ferromagnetic (FM) kagome lattice [1, 2]. In such a system, the nearest-neighbor hopping of the tight-binding model naturally generates Dirac cones at the K points of the Brillouin zone (BZ) corners. These Dirac cones are further split into spin-up and spin-down channels by FM exchange. When an out-of-plane magnetization and Kane-Mele-type spin-orbit coupling (SOC) are introduced, a Chern gap opens at the spin-polarized Dirac crossing, giving rise to Chern-gapped Dirac fermions. This theoretical framework was experimentally confirmed in the kagome magnet $TbMn_6Sn_6$ [3], where scanning tunneling microscopy measurements revealed an electronic structure fully consistent with the Chern-gapped Dirac model. Crucially, the pristine Mn kagome lattice and the strong out-of-plane magnetization in $TbMn_6Sn_6$ are indispensable for stabilizing this exotic topological state. Thus, engineering the symmetry of the kagome lattice—and consequently, its spin dynamics—offers a promising pathway to uncover new topological phenomena.

In the hexagonal $R$Mn$_6$Sn$_6$ family ($R$ = rare earth), magnetism exhibits a strong dependence on the $R$-site ion [4–11]. Systems with nonmagnetic $R$ ions feature strong FM intralayer Mn-Mn interactions with weak easy-plane anisotropy, while competing FM and antiferromagnetic (AFM) interlayer couplings lead to helical magnetism and a topological Hall effect (THE) [4, 7, 12, 13]. In contrast, when $R$ is moment-bearing, the magnetic behavior is dictated by the $R$-ion anisotropy and AFM $R$-Mn coupling [14]. For instance, $TbMn_6Sn_6$ exhibits a unique uniaxial collinear FM state, stabilized by the strong uniaxial anisotropy of $Tb^{3+}$ ions and AFM Tb-Mn interactions. However, $R$Mn$_6$Sn$_6$ compounds are prone to magnetic instabilities, undergoing spin reorientations under temperature variations or external magnetic fields. In $TbMn_6Sn_6$, for example, the ferrimagnetic state transitions from out-of-plane to easy-plane collinear magnetization above 310 K [11, 15]. Moreover, applying a magnetic field along the hard axis can induce a first-order spin reorientation [16].

These observations highlight that tuning the rare-earth ion provides a viable strategy to manipulate magnetic instabilities, offering a potential route to switch between different topological magnetic phases [17]. Alternatively, engineering the Mn kagome layer—by modifying its structural symmetry and exchange interactions—could similarly enable control over topological states. Yet, this promising avenue remains experimentally unexplored.

In this study, we strategically designed $TbCr_2Mn_4Sn_6$, where nonmagnetic Cr substitution serves a dual purpose to isolate Tb ions and weaken both the AFM Tb-Mn and FM intralayer Mn-Mn interactions. This targeted engineering enables precise manipulation of spin dynamics and, consequently, topological properties. Key to this approach is the preferential concentration of Cr atoms in the single $Mn_1$ layer, which disrupts the kagome lattice symmetry and effectively decouples the Tb and Mn magnetic networks. This symmetry-breaking yields a remarkable plateau-like THE, persisting across a broad temperature range (150–220 K). The phenomenon originates from the contrasting exchange stiffness between the two Mn layers, where the $Mn_1$ layer, with its weakened in-plane exchange coupling, achieves full magnetization readily, while the $Mn_2$ layer retains stronger exchange interactions. This disparity drives the formation of distinct magnetic domains in each Mn layer under applied fields, culminating in an abrupt magnetic domain transition at 1 T. The transition is marked by a striking plateau in the topological Hall resistivity, reflecting the interplay between layer-specific magnetism and emergent topology.

For comprehensive experimental and theoretical details—including single-crystal growth, structural characterization (XRD, EDX), magnetic measurements, MFM imaging, electronic transport studies, and supporting first-principles calculations—see in the Supplemental Material (SI).

**Results and Discussion.**

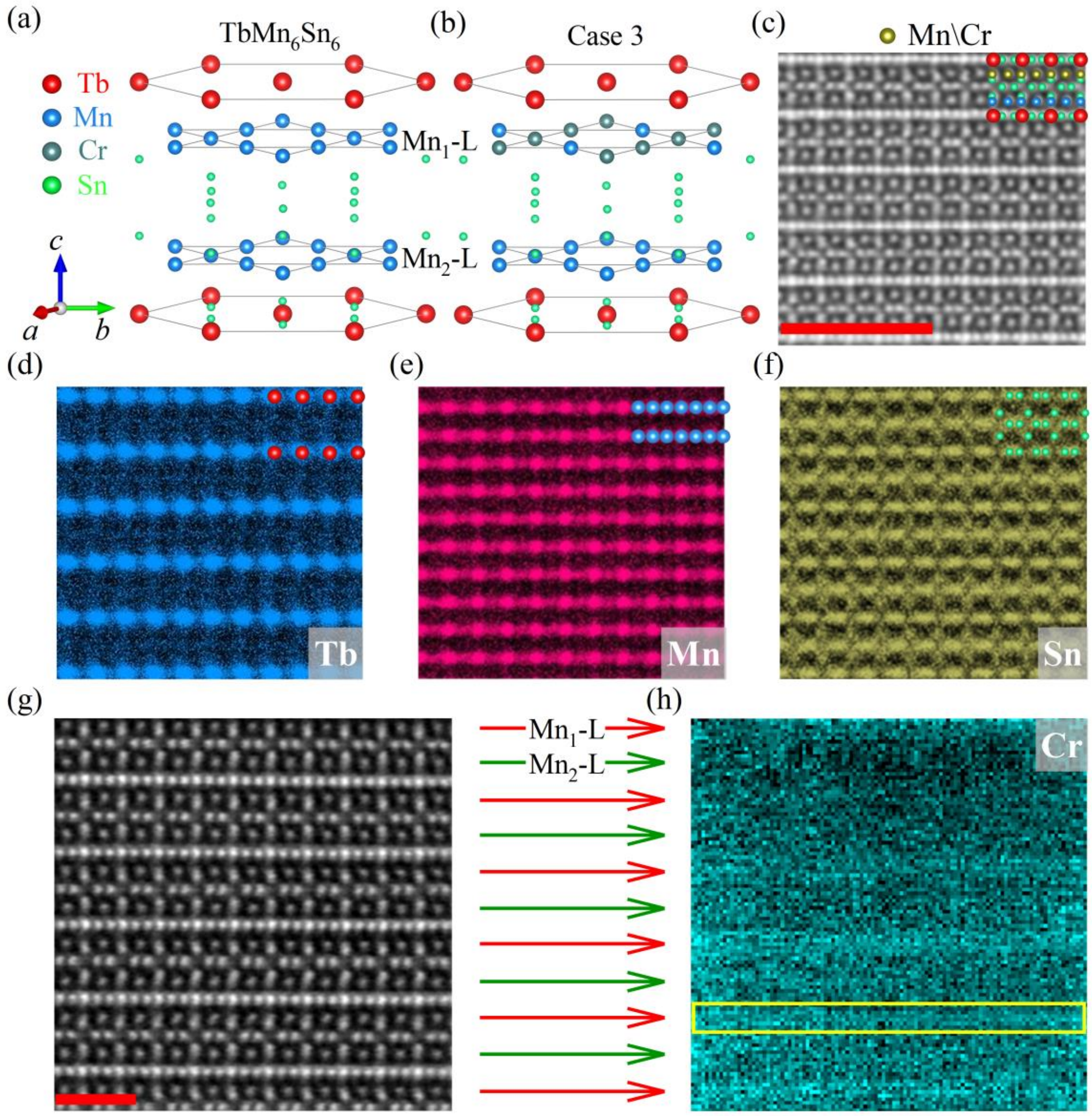

**FIG. 1. Structural characterization of $TbMn_6Sn_6$ and Cr-substituted $TbCr_2Mn_4Sn_6$ crystals.** (a) Crystal structure schematic of pristine $TbMn_6Sn_6$, showing the ideal kagome lattice arrangement. (b) Modified crystal structure of $TbCr_2Mn_4Sn_6$, highlighting the selective Cr substitution in the $Mn_1$ kagome layer (indicated by purple spheres). (c) HAADF-STEM image revealing the atomic interlayer stacking in $TbCr_2Mn_4Sn_6$. (d-f) Atomic-resolution EDS elemental maps of (d) Tb (red), (e) Sn (green), and (f) Mn (blue) within the (110) crystallographic plane (scale bar: 2.5 nm). (g) High-magnification HAADF-STEM image (scale bar: 1 nm) and (h) corresponding Cr elemental map, confirming the preferential occupation of Cr atoms in the $Mn_1$ sites. Color scheme: Tb (red spheres), Mn (blue spheres), Sn (green spheres), Cr (dark cyan spheres).

As illustrated in Fig. 1(a), $TbMn_6Sn_6$ crystallizes in a hexagonal structure (space group *P*6/*mmm*), comprising alternating layers of Tb and two pristine Mn kagome lattices within each unit cell [3]. The Mn atoms occupy two distinct sites, denoted as $Mn_1$ and $Mn_2$, corresponding to the two neighboring kagome layers. To engineer the electronic and magnetic properties of the Mn kagome lattice, we introduced Cr substitution for Mn. Determining the precise Cr occupancy is critical, with three possible configurations explored (Figs. S1(a)-S1(c), SI). First-principles calculations on a large supercell (Fig. S2) revealed that Cr preferentially substitutes into a single $Mn_1$ layer, as this configuration is energetically favored by approximately 26.9–83.7 meV per formula unit over metastable alternatives (Fig. 1(b)). This selective occupation yields an average magnetic moment of 1.88 $\mu_B$ per Mn/Cr atom, in excellent agreement with experimental measurements (Table 1) [18].

**Table 1**. The calculated magnetic moments of Tb, Mn and Cr atoms. The spins of Tb and Mn stay in the AFM coupling.

| Atom | Magnetic moment |
| --- | --- |
| Cr | 0.54$\mu_B$ |
| Cr | 0.54$\mu_B$ |
| Mn ($Mn_1$ layer) | 2.82$\mu_B$ |
| Mn ($Mn_2$ layer) | 2.47$\mu_B$ |
| Mn ($Mn_2$ layer) | 2.43$\mu_B$ |
| Mn ($Mn_2$ layer) | 2.48$\mu_B$ |
| Tb | -8.64$\mu_B$ |

Chemical analysis confirmed the stoichiometric purity of the synthesized crystals (Figs. S3(a)-S3(b), SI). Powder X-ray diffraction (PXRD) at room temperature verified the hexagonal structure of $TbCr_2Mn_4Sn_6$, with all peaks indexed accordingly (Fig. S3(c), SI). The high crystalline quality was further evidenced by the sharp (004) Bragg peak (FWHM = 0.034°, Fig. S4(d), SI). Atomic-resolution HAADF-STEM imaging along the (110) plane (Fig. 1(c)) confirmed the expected crystal structure, while energy-dispersive X-ray spectroscopy (EDS) mapping (Figs. 1(d)-1(f)) spatially resolved the distributions of Tb, Mn, and Sn. Crucially, layer-resolved Cr mapping (Fig. 1(g)) demonstrated that Cr dopants predominantly occupy the $Mn_1$ layer, aligning with DFT predictions (Fig. 1(h)). These results collectively establish that Cr selectively substitutes the $Mn_1$ kagome layer, breaking the pristine lattice symmetry.

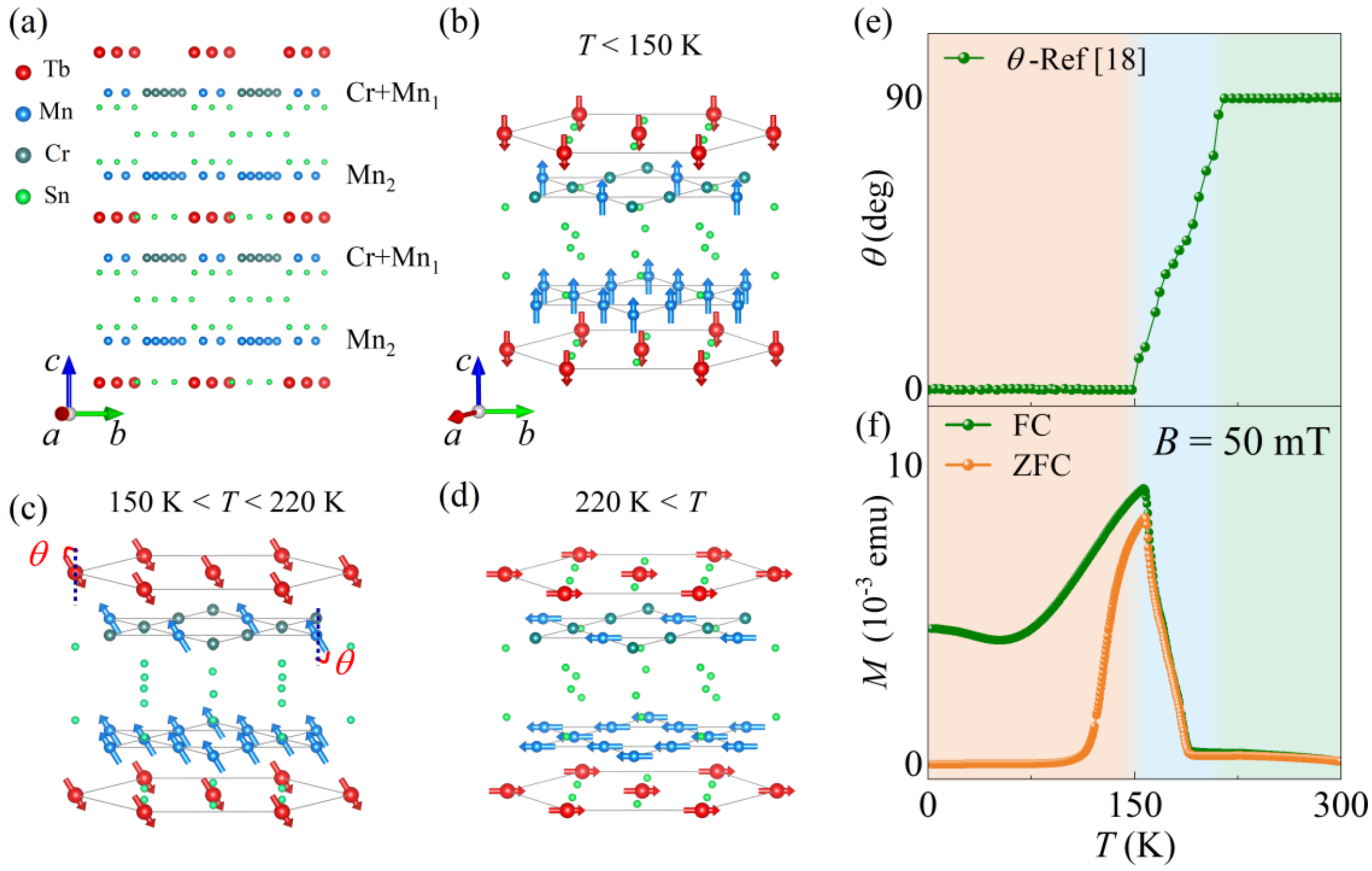

**FIG. 2. Crystal and magnetic structure evolution of $TbCr_2Mn_4Sn_6$.** (a) Crystal structure of $TbCr_2Mn_4Sn_6$, highlighting the Cr-substituted kagome layers. Magnetic Structures of $TbCr_2Mn_4Sn_6$ at (b) $T \leq 150$ K, (c) 150 K $\leq T \leq$ 220 K and (d) 220 K $\leq T$. (e) Temperature evolution of the canting angle ($\theta$) between Tb and Mn/Cr sublattice moments relative to the *c*-axis, adapted from Ref. [18]. (f) Temperature-dependent magnetization measured at 50 mT (*B*//*c*-axis), revealing distinct magnetic transitions that correlate with the structural changes shown in (b-d).

The crystal structure of $TbCr_2Mn_4Sn_6$ (Fig. 2(a)) consists of a hexagonal Tb layer and a kagome ($Cr/Mn_1$) layer, stacked along the *c*-axis in the sequence of $Mn_2$-Tb-($Cr/Mn_1$)-$Mn_2$-Tb-($Cr/Mn_1$) [18]. Neutron diffraction studies [18] reveal its complex magnetic structures. Below 150 K (Fig. 2(b)), the Tb and Mn sublattices exhibit antiferromagnetic (AFM) coupling along the *c*-axis, while above 220 K (Fig. 2(d)), the spins reorient into the basal plane, while retaining AFM Tb-Mn coupling. A key unifying feature is the robust in-plane ferromagnetic (FM) alignment of Mn spins, which remain antiparallel to Tb moments due to strong Tb-Mn AFM exchange. This hierarchical spin arrangement underscores the interplay between local symmetry breaking (Cr doping) and emergent magnetism.

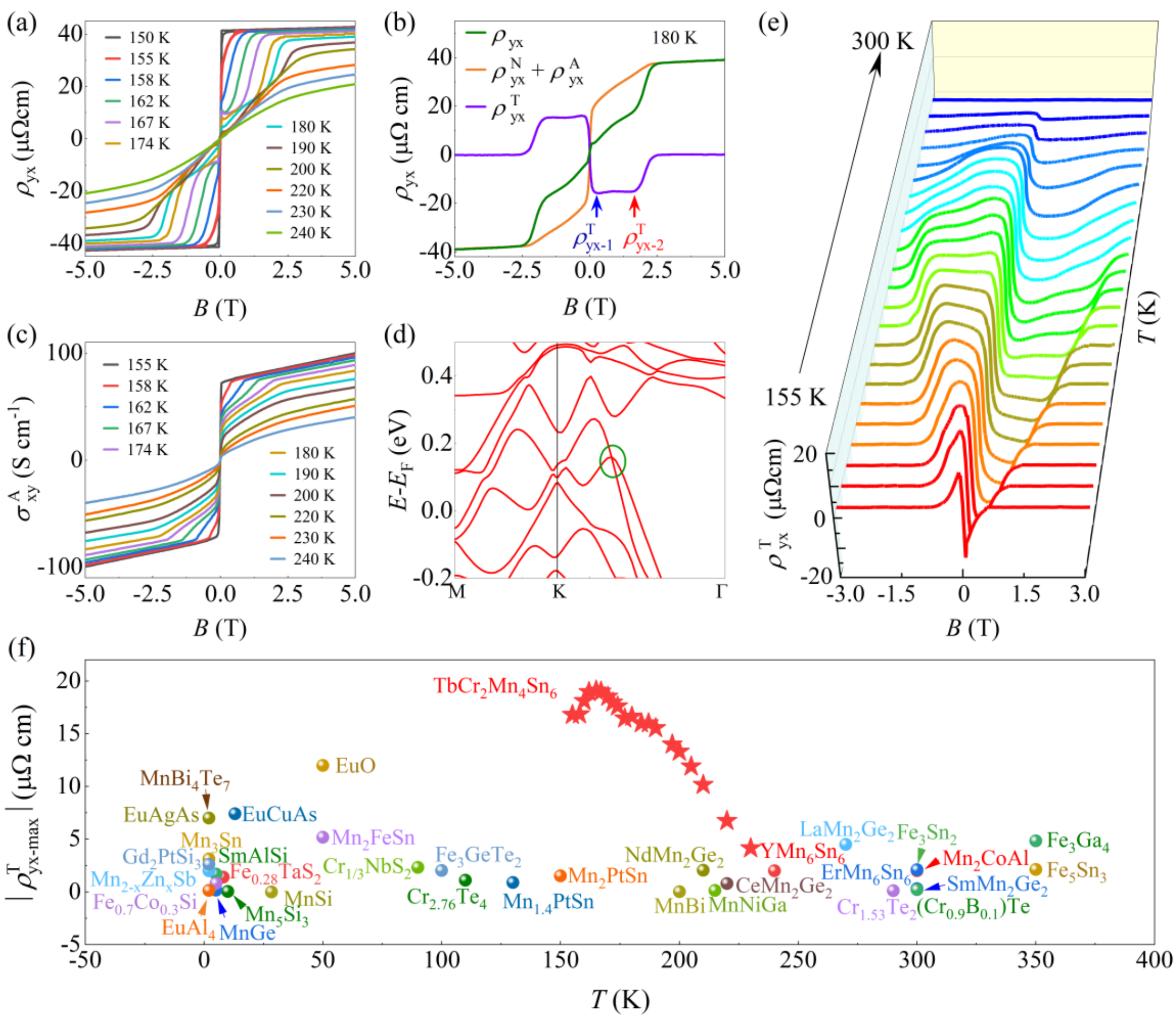

**FIG. 3. Magnetotransport properties and electronic structure of $TbCr_2Mn_4Sn_6$.** (a) Temperature evolution of $\rho_{xy}$ under out-of-plane magnetic field configuration ($B$ // $c$-axis), demonstrating the emergence of anomalous Hall effect below 150 K. (b) Deconvolution of the 180 K Hall response into normal (orange, $R_0B + 4\pi R_S M$) and topological (purple, $\rho^{T}_{yx}$) components, revealing significant nonlinear contributions. (c) Temperature-dependent $\sigma^{A}_{xy}$ showing systematic variation with magnetic field. (d) First-principles band structure calculations along the high-symmetry M-K-Γ direction, with green circles highlighting the accidental band crossings responsible for the observed anomalous Hall effect. (e) Temperature progression of $\rho^{T}_{yx}$, exhibiting characteristic plateau formation between 150-220 K. (f) Comparative analysis of maximum $|\rho^{T}_{yx}|$ values with reported topological magnets (Refs. [22-55]), where pentagrams denote the record-breaking values achieved in this work across different temperatures.

Similar to its parent compound $TbMn_6Sn_6$, $TbCr_2Mn_4Sn_6$ exhibits a pronounced spin reorientation transition—but with a striking shift in temperature regime. While $TbMn_6Sn_6$ undergoes this transition only above 290 K [19, 20, 21], $TbCr_2Mn_4Sn_6$ displays analogous behavior within a significantly lower range (150–220 K), as depicted in Fig. 2(c). During this transition, the spins of Tb and Mn progressively deviate from the *c*-axis by an angle $\theta$, eventually reorienting into the basal plane (Fig. 2(c)). The temperature evolution of $\theta$, plotted in Fig. 2(e) (data from Ref. [18]), captures this continuous rotation, confirming the out-of-plane to in-plane spin reconfiguration. Critically, the magnetization data in Fig. 2(f) exhibit a transition temperature that aligns precisely with the $\theta$(T) behavior, providing independent validation of the spin reorientation mechanism. This consistency between structural ($\theta$) and magnetic ($M$) responses underscores the robustness of the observed phenomenon.

To elucidate the correlation between magnetism and magnetotransport behavior, we conducted systematic measurements of the Hall resistivity ($\rho_{yx}$) under varying temperatures and magnetic fields. Fig. S5 presents the temperature and field dependence of $\rho_{yx}$ for Sample 1 between 2 K and 300 K. Strikingly, the anomalous Hall effect (AHE) emerges below 150 K, while the hysteresis loop of $\rho_{yx}$ gradually diminishes as the temperature increases from 150 K to 220 K. Concurrently, multiple kinks emerge in low magnetic fields, exhibiting a remarkable correspondence with the isothermal magnetization features observed in Sample 2 (Fig. S6).

To unravel the origin of these kinks, we performed detailed measurements of $\rho_{yx}$ and magnetization $M$(H) for Sample 3 within the 150–220 K range (Fig. 3 and Fig. S7). Fig. 3(a) clearly depicts the temperature evolution of these kinks. The total Hall resistivity can be expressed as $\rho_{yx} = \rho^{N}_{yx} + \rho^{A}_{yx} + \rho^{T}_{yx} = R_H B + 4\pi R_S M + \rho^{T}_{yx}$, where $\rho^{N}_{yx}$ arises from the normal Hall effect, $\rho^{A}_{yx}$ represents the AHE, and $\rho^{T}_{yx}$ originates from the topological Hall effect (THE), attributed to noncollinear spin textures hosting finite spin chirality [56]. For clarity, Fig. 3(b) compares the field-dependent $\rho_{yx}$, $\rho^{N}_{yx}+\rho^{A}_{yx}$, and $\rho^{T}_{yx}$ at 180 K, unambiguously revealing a nonzero $\rho^{T}_{yx}$. Intriguingly, the THE in $TbCr_2MnSn_6$ exhibits two distinct peaks ($\rho^{T}_{yx\text{-}1}$ and $\rho^{T}_{yx\text{-}2}$, labeled in Fig. 3(b)) of comparable magnitude, yielding a plateau-like behavior—a feature distinct from those reported in other systems [23, 44–48, 57–59]. Fig. 3(c) displays the temperature-dependent anomalous Hall conductivity (AHC, $\sigma^{T}_{xy}$). At 150 K, the maximum $\rho^{A}_{yx}$ and $\sigma^{A}_{xy}$ reach 42.75 μΩ cm and 99.5 $\Omega^{-1}cm^{-1}$, respectively.

Bulk $TbMn_6Sn_6$ possesses $D_{6h}$ point group symmetry, which safeguards Dirac electronic states [3]. However, Cr substitution in $TbCr_2Mn_4Sn_6$ disrupts the kagome lattice of the $Mn_1$ layer, reducing the symmetry to $C_2$ and altering the crystal field near the $Mn_2$ kagome layer. Consequently, the Dirac point at the K point—protected by $D_{6h}$ symmetry—is lifted, and the AHE in $TbCr_2Mn_4Sn_6$ primarily stems from accidental band crossings [19], as corroborated by theoretical calculations (Fig. 3(d)). The temperature evolution of $\rho^{T}_{yx}$ (Fig. 3(e)) further reveals a plateau-like behavior, while Fig. 3(f) benchmarks the maximal $|\rho^{T}_{yx}|$ values reported across various systems. Notably, $TbCr_2Mn_4Sn_6$ exhibits a record-high $|\rho^{T}_{yx}|$ amplitude of 19.1 μΩ cm, surpassing all known magnetic materials [22–55].

We now turn to the origin of the striking plateau-like behavior observed in $\rho^{T}_{yx}$. In $TbMn_6Sn_6$, non-coplanar spin textures of Tb, $Mn_1$, and $Mn_2$ atoms stabilize topological magnetic skyrmions near room temperature, giving rise to the topological Hall effect (THE) [49]. For $TbCr_2Mn_4Sn_6$, Fig. 4(a) presents the phase diagram of $\rho^{T}_{xy}$, while Fig. 4(b) displays the magnetic field dependence of $\rho^{T}_{yx}$ at 180 K. Notably, the inset reveals an enlarged view of the low-field regime, where two distinct peaks flank a pronounced minimum at 0.95 T—a hallmark of the unconventional plateau-like THE.

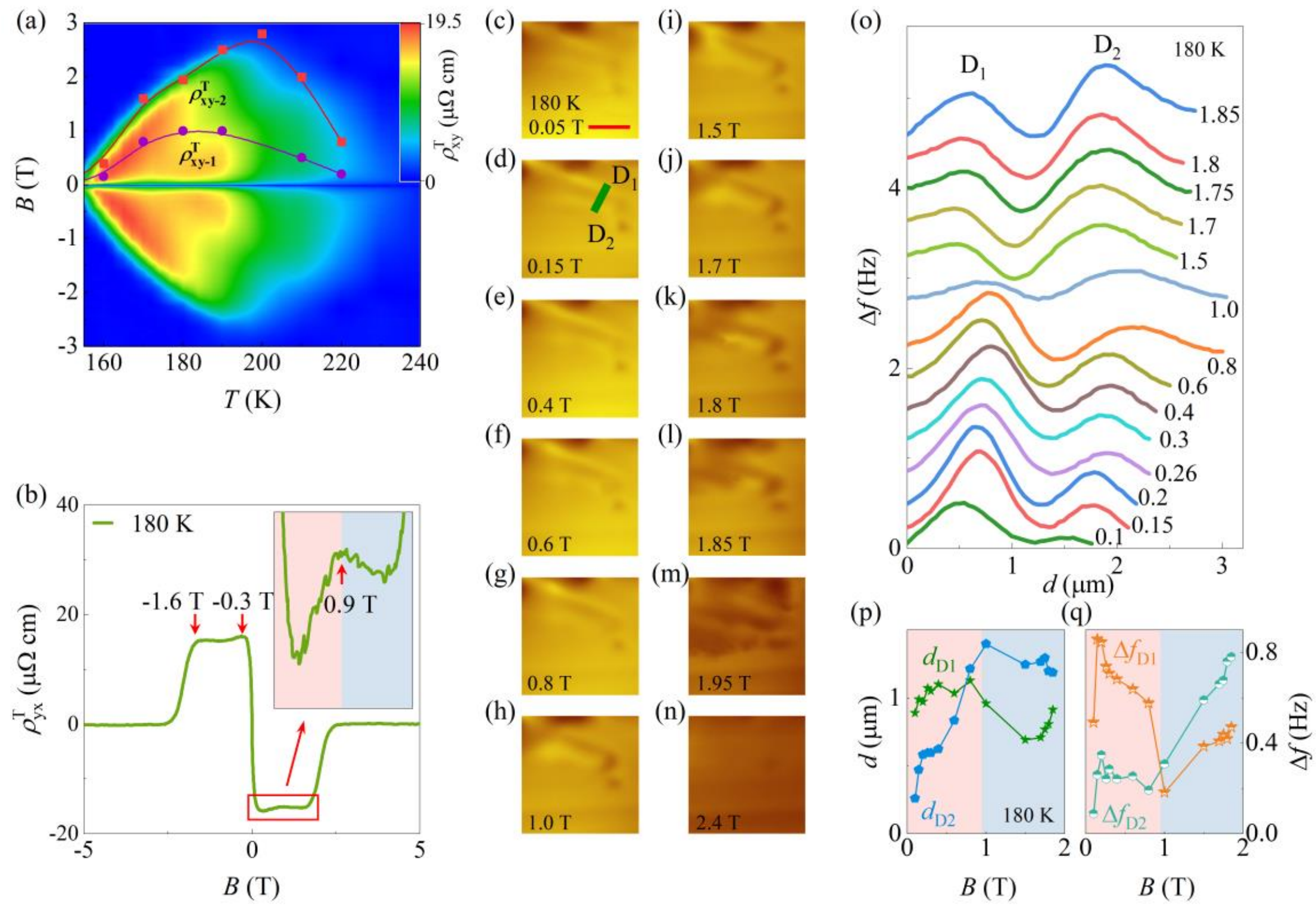

**FIG. 4. Interplay between topological Hall response and magnetic domain evolution in $TbCr_2Mn_4Sn_6$.** (a) Phase diagram of $\rho^{T}{}_{xy}$ as a function of temperature and magnetic field. The purple dashed line marks the onset field for domain wall motion ($D_1/D_2$ width variation), while the red dashed line indicates the critical field for complete domain annihilation. (b) Field-dependent $\rho^{T}{}_{yx}$ at 180 K, with inset highlighting the low-field regime (0.2-1.8 T) where domain reconfiguration occurs. (c-i) Real-space magnetic domain evolution visualized by MFM at 180 K under increasing perpendicular fields (0.05-2.4 T, scale bar: 2 $\mu$m in c). The stripe domains $D_1/D_2$ exhibit progressive modification leading to eventual disappearance. (o) Quantitative MFM line profiles (along blue trace in d) demonstrating field-dependent domain wall dynamics. (p) Systematic variation of $D_1/D_2$ domain widths with applied field, showing abrupt changes near 1 T that correlate with the $\rho^{T}{}_{yx}$ features in (b). (q) Corresponding evolution of frequency shift ($\Delta f$) for both domains, providing direct evidence for their distinct magnetic stiffness.

To decipher the microscopic mechanism underlying this giant THE, we performed magnetic force microscopy (MFM) measurements on $TbCr_2Mn_4Sn_6$ single crystals. Figs. 4(c)–4(n) depict the evolution of magnetic domains under increasing magnetic fields (0.05–2.4 T), imaged in the same region (scale bar: 2 $\mu$m, Fig. 4(c)). Two distinct stripe domains, $D_1$ and $D_2$, dominate the landscape at low fields (Fig. 4(c)). Fig. 4(o) tracks the MFM signal profile along the blue line in Fig. 4(d), while Figs. 4(p) and 4(q) quantify the field-dependent variations in domain width and intensity for $D_1$ and $D_2$, respectively (extracted via the schematic in Fig. S14). Strikingly, both domains undergo a sharp transition near 1 T, mirroring the THE anomaly at 0.95 T in $\rho^{T}_{yx}$ (Fig. 4(b)). Further evidence of domain-THE correlation emerges from field-polarity-independent behavior. Figs. S15(a)–S15(d) demonstrate that domain evolution under positive (0 → 2.45 T) and negative (0 → −2.5 T) fields is nearly identical, confirming reversibility. This domain dynamics persists up to 220 K, paralleling the THE's thermal stability. Complementary MFM studies across 150–220 K (Fig. S16) reveal critical fields for domain transitions (circles) and collapse (squares), whose temperature dependence aligns remarkably with the extremal points ($\rho^{T}_{yx\text{-}1}$ and $\rho^{T}_{yx\text{-}2}$) in the THE (Fig. 4(a)).

The intimate correlation between magnetic domain restructuring and THE anomalies unambiguously establishes that the plateau-like $\rho^{T}_{yx}$ arises from the field-driven evolution of domains $D_1$ and $D_2$. This mechanism, distinct from conventional skyrmion-driven THE, underscores the rich interplay between real-space spin textures and topological transport in $TbCr_2Mn_4Sn_6$.

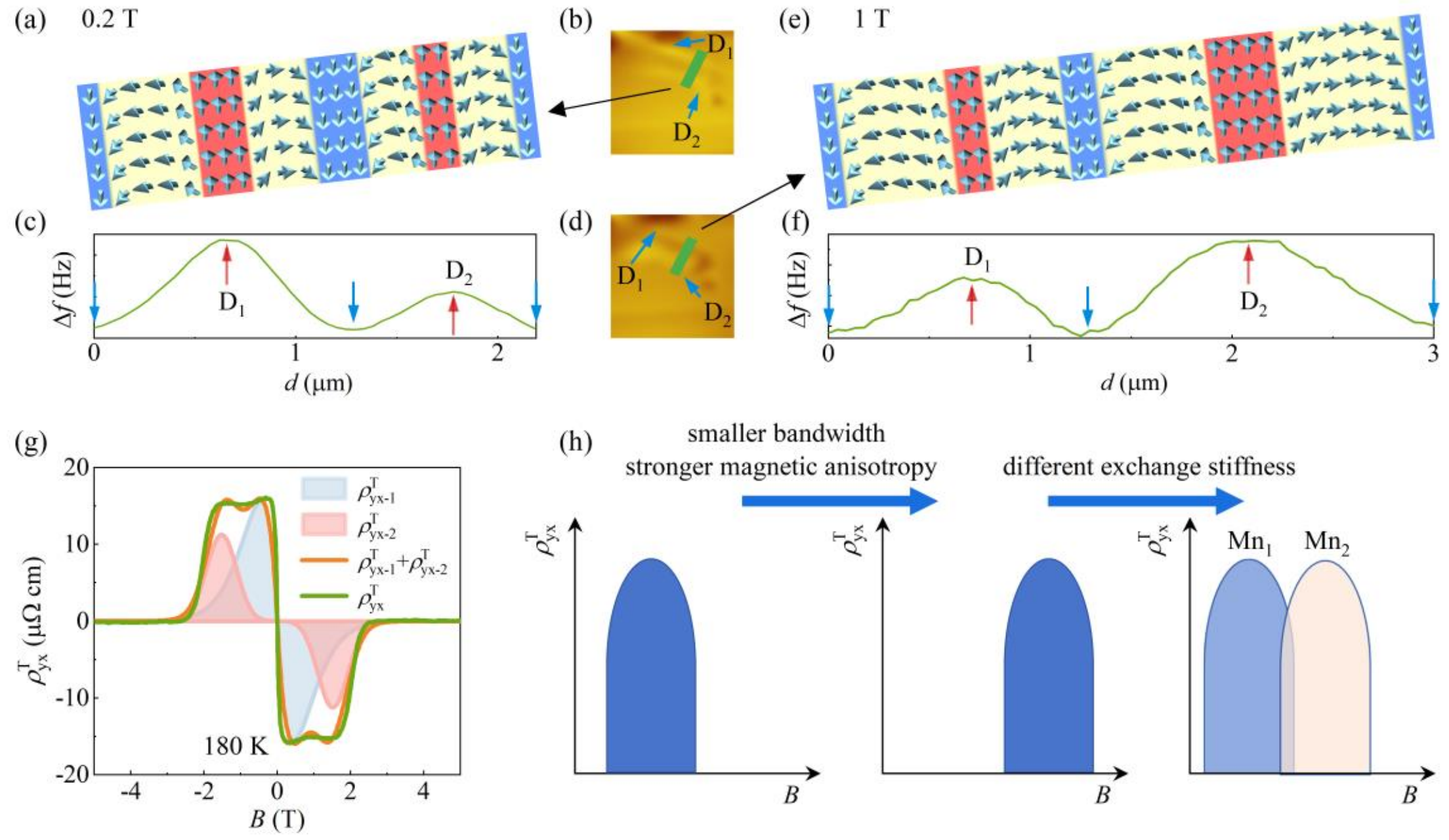

**FIG. 5. Microscopic origin of the plateau-like topological Hall effect in $TbCr_2Mn_4Sn_6$.** (a, e) Schematic spin configurations along the blue lines in MFM images (b, d) at 180 K under 0.2 T and 1 T, respectively, showing the evolution from stripe domains to a partially polarized state. Red and blue arrows indicate spin-up and spin-down orientations, illustrating the non-collinear spin texture. (c, f) Corresponding MFM signal line profiles reveal distinct domain structures—narrow, well-defined walls at 0.2 T (c) versus broadened, reoriented domains at 1 T (f). (g) Decomposition of $\rho^{T}_{yx}$ (grass green) into two components ($\rho^{T}_{yx-1}$, light blue; $\rho^{T}_{yx-2}$, pink) via peak fitting, with their sum (orange) reproducing the plateau-like feature. (h) Proposed mechanism: Cr doping reduces bandwidth, enhancing magnetic anisotropy in the $Mn_2$ layer and raising its critical domain-switching field, while weakened exchange stiffness in the $Mn_1$ layer lowers its domain stability. This disparity creates dual magnetic domains ($D_1$/$D_2$), whose sequential polarization under increasing field produces the step-like $\rho^{T}_{yx}$ plateau—a direct signature of layer-resolved topological spin textures.

The magnetic domains undergo a pronounced transition near 1 T as the external field varies. Figs. 5(a) and 5(e) present cross-sectional spin configurations at 0.2 T and 1 T, respectively, corresponding to the MFM signals shown in Figs. 5(c) and 5(f). These configurations reveal a marked transformation in domain $D_2$'s structure. The MFM signals, acquired along the blue trajectories indicated in Figs. 5(b) and 5(d), further corroborate this transition. At 180 K, $\rho^{T}_{yx}$ distinctly resolves into two characteristic peaks $\rho^{T}_{yx-1}$ and $\rho^{T}_{yx-2}$) as shown in Fig. 5(g), demonstrating the composite nature of this phenomenon. This dual-peak structure naturally leads us to investigate the origin of the corresponding dual magnetic domains in $TbCr_2Mn_4Sn_6$ under applied magnetic fields.

The underlying mechanism becomes clear when examining the electronic structure modifications induced by Cr doping. As illustrated in Fig. 3(d), Cr substitution destroys the dispersionless Dirac cone characteristic of the parent compound, while simultaneously reducing the itinerant electron bandwidths significantly. Perturbation theory provides crucial insight. The $n$-order magnetic anisotropy constant ($K_n$) for $3d$-electrons follows $K_n \sim \lambda^{2n}/W^{2n-1}$, where $\lambda$ represents the SOC strength and $W$ denotes the bandwidth [60]. This relationship reveals two important consequences of bandwidth reduction. For the $Mn_2$ layer, the enhanced magnetic anisotropy makes spin reorientation more difficult, thereby increasing the critical field for domain formation compared to $TbMn_6Sn_6$ [20], while in the $Mn_1$ layer, the presence of nonmagnetic Cr atoms dramatically reduces the spin exchange stiffness, leading to significantly lower critical fields for domain formation. This stark contrast in exchange stiffness between the $Mn_1$ and $Mn_2$ layers, induced by Cr doping, creates the necessary conditions for dual magnetic domain formation. The resulting domain structure, schematically illustrated in Fig. 5(h), directly accounts for both the observed plateau-like topological Hall effect and its characteristic two-peak structure in $\rho^T{}_{yx}$.

In conclusion, through strategic Cr substitution in the pristine Mn kagome layer of $TbMn_6Sn_6$, we have engineered a remarkable transformation in both structural symmetry and magnetic properties. This substitution achieves three critical modifications: (i) It breaks the original $D_{6h}$ point group symmetry, reducing it to $C_2$ symmetry, (ii) It disrupts the collinear spin configuration under applied magnetic fields, and (iii) it creates a unique spin texture that manifests in two distinct electronic transport phenomena. Below 150 K, the system exhibits an anomalous Hall effect arising from accidental crossings in the reconstructed electronic band structure. More strikingly, between 150-220 K, we observe an unprecedented bulk topological Hall effect characterized by a distinctive plateau-like resistivity - the largest reported to date in any magnetic material. Our MFM investigations reveal the microscopic origin of these phenomena, where the Cr-induced modification creates a dramatic disparity in exchange coupling between the $Mn_1$ and $Mn_2$ layers. While the $Mn_1$ layer, with its weakened exchange interaction, readily magnetizes under applied fields, the $Mn_2$ layer maintains stronger exchange interaction. This imbalance produces competing magnetic domains that undergo an abrupt transition near 1 T - precisely where the topological Hall resistivity develops its plateau feature.

The perfect correlation between the temperature- and field-evolution of the topological Hall effect and the corresponding domain reconfiguration demonstrates that the plateau behavior directly results from the dual-domain structure, the domain formation is governed by the layer-dependent exchange stiffness, and the giant topological Hall effect emerges from this engineered spin texture. These findings establish kagome layer engineering as a powerful strategy for creating exotic topological states in magnetic kagome materials. By selectively tuning exchange interactions in different atomic layers, we can design complex spin architectures that give rise to novel transport phenomena, opening new avenues for exploring and controlling topological quantum materials.

## Methods

### Single crystal growth and characterizations

High-quality $TbCr_2Mn_4Sn_6$ single crystals were synthesized via a self-flux method using ultra-high purity elements: Tb ingot (99.99%), Cr powder (99.95%), Mn block (99.9%), and Sn granules (99.999%). The constituent elements were combined in a molar ratio of 1: 2: 4: 20 and sealed in an alumina crucible under vacuum within a quartz ampoule. The growth process employed a precisely controlled thermal profile including heating to 1100°C over 15 hours with a 20-hour homogenization period, primary crystallization at 900°C for 10 hours following a 5-hour cooling interval, controlled cooling to 500°C at 1.5°C/h to optimize crystal quality, rapid flux removal via high-speed centrifugation at 500°C and a final cooling to room temperature in ambient conditions. This protocol yielded millimeter-scale single crystals exhibiting characteristic black metallic luster. Comprehensive structural characterization was performed using powder X-ray diffraction (PXRD) with Cu $K_\alpha$ radiation ($\lambda$ = 1.5418 Å) on the (001) crystallographic plane, single-crystal X-ray diffraction (SXRD) employing Mo $K_{\alpha 1}$ radiation ($\lambda$ = 0.71073 Å) at 298 K, and energy-dispersive X-ray spectroscopy (EDS) for stoichiometric verification.

### Magnetization and electrical transport measurements

The magnetic properties were systematically investigated using a Quantum Design MPMS-3 superconducting quantum interference device (SQUID) magnetometer. Both temperature-dependent magnetic susceptibility and field-dependent isothermal magnetization measurements were performed to comprehensively characterize the magnetic behavior. For transport measurements, we employed a Quantum Design PPMS DynaCool system with a standard six-wire configuration to ensure precise determination of both longitudinal resistivity and Hall effect. This approach effectively eliminates potential contact misalignment artifacts. The Hall resistivity measurements revealed a distinctive double-peak structure, with well-defined features at $\rho^{T}_{yx-1}$ and $\rho^{T}_{yx-2}$. Detailed quantitative analysis of the topological Hall response was performed at the characteristic temperature of 180 K. The data were analyzed using a peak deconvolution methodology, implemented through Gaussian multi-peak fitting in Origin software. This rigorous analysis protocol enabled precise determination of the individual contributions to the total Hall response, providing crucial insight into the underlying physical mechanisms.

**Cs-corrected scanning transmission electron microscopy Measurements**

The double Cs-corrected scanning transmission electron microscopy (STEM, Grand JEM-ARM300F) was used to analyze the atomic structure. The microscopy is equipped with a cold field emission gun and operated at the accelerating voltage of 300 kV. The used crystal was cut by a focused ion beam (FIB, Grand JIB-4700F) technique along the (110) plane.

**Magnetic fore microscopy measurements**

We conducted nanoscale magnetic structure imaging using a custom-built low-temperature magnetic force microscopy (MFM) system integrated with a 12 T superconducting magnet [61]. The measurements employed commercial piezoresistive silicon probes (PRSA-L300-F50-STD from SCL-Sensor.Tech) featuring a 40 nm thick ferromagnetic Co coating, with a room-temperature resistance of 2.1 kΩ and resonance frequency of 35 kHz. Prior to measurements, the probes were magnetized using a neodymium magnet, exhibiting a coercivity of ~250 Oe at 5 K. The system was controlled using an RHK Technology R9 controller equipped with a phase lock loop. For optimal imaging quality, we first prepared the sample surface by fabricating a smooth 70 × 70 $\mu m^2$ area on the single crystal bulk sample using focused ion beam-scanning electron microscopy (FIB-SEM). The probe tip was carefully positioned on the target area under optical microscopy, with the tip-sample distance minimized at room temperature to prevent temperature-induced drift during cooling. Our imaging protocol involved a two-scan procedure: first acquiring topographical information via tapping mode, followed by magnetic structure imaging in lift mode with the tip raised 150 nm above the surface. During MFM operation, the cantilever's resonant frequency variations directly corresponded to the out-of-plane stray magnetic field gradient [62], where dark and bright contrast regions indicated attractive (parallel magnetization) and repulsive (antiparallel magnetization) tip-sample interactions relative to the applied external field, respectively. This comprehensive approach enabled high-resolution visualization of magnetic domain structures under extreme conditions.

**First-principles calculations**

The first-principles calculations were carried out in the framework of the generalized gradient approximation (GGA) functional [63] of the density functional theory with projector augmented wave method [64] implemented in the Vienna ab initio simulation package (VASP) [65]. In the calculations, the on-site Hubbard parameter U = 7 eV is applied on Tb-f states [66]. Then the Bloch states are projected to the Wannier functions [67] to build the tight-binding Hamiltonian, and we then used WannierTools package [68] to calculate the anomalous Hall conductivity.

## Supporting Information

Supplementary information is available for this paper at https://xxxx.

## Data availability

The data that support the findings of this study are available from the corresponding authors upon reasonable request.

## Competing interests

The authors declare no competing interests.

**Acknowledgements**

The authors acknowledge the National Key R&D Program of China (Grants No. 2024YFA1408400) and the National Nature Science Foundation of China (Grant No. 11934017). Y.F.G. acknowledges the open research fund of Beijing National Laboratory for Condensed Matter Physics (2023BNLCMPKF002). X.W.F. acknowledges National Key R&D Program of China (Grants No. 2020YFA0309300), the National Natural Science Foundation of China (Grant No. 12127803), and the "Fundamental Research Funds for the Central Universities", Nankai University (Grant No. 63243194, 9242000728). W.X. thanks the National Natural Science Foundation of China (Grants No. 12404186) and the Shanghai Sailing Program (23YF1426900). S.H.Z. was supported by the National Key Research and Development Program of China (No. 2024YFA1410300), the National Natural Science Foundation of China (No. 12304217), the Natural Science Foundation of Hunan Province (No. 2025JJ60002) and the Fundamental Research Funds for the Central Universities from China (No. 531119200247). Q.Y.L. thanks the National Key R&D Program of China (Grant 2023YFA1607701) and National Natural Science Foundation of China (Grants 51627901, U1932216). We thank the staff members of the SMA System (https://cstr.cn/31125.02.SHMFF.SM2.SMA) at the Steady High Magnetic Field Facility, CAS (https://cstr.cn/31125.02.SHMFF) for providing technical support and assistance in data collection and analysis. We also thank Prof. Jinguang Cheng for the help with magnetization measurements.

**Author Contributions.**

Y.F.G. and W.X. conceived the project; W.X. synthesized the single crystals and carried out structural characterizations, measured the magnetotransport properties and analyzed the data with the help from J.Y., J.W.L., Y.R.W., H.D., X.Q.L., J.T.G., Y.X.L., M.C., J.L.J., X.W., X.R.L., W.B.W., Y.L.C., X.W.F., R.D.Z.; H.N.W, K.Q and Z.Z.Y.

measured the atomic-resolution crystal structure by using the Cs-corrected HAADF-STEM; A.L.W., W.J.M., Y.B.H. and Q.Y.L. performed the MFM measurements; W.X., A.L.W., W.J.M. analyzed the MFM data; S.H.Z. performed the first-principles calculations; W.X., A.L.W. and J.Y. contributed equally to this work; W.X., S.H.Z., Z.Z.Y. and Y.F.G. wrote the manuscript with input from all authors.

# SI

## Supplementary Note 1: Structural Characterizations of $TbCr_2Mn_4Sn_6$

Our comprehensive investigation combining first-principles calculations and experimental characterization provides definitive evidence for Cr substitution behavior in $TbCr_2Mn_4Sn_6$. Theoretical analysis of three possible substitution configurations (Fig. S1) reveals Case 3 - where Cr atoms occupy sites within the same Mn layer - as the most energetically favorable, being 201.73 meV and 179.39 meV per formula unit more stable than Cases 1 and 2 respectively. This preference is further confirmed by extended supercell calculations (Fig. S2), which show metastable configurations with Cr distributed across neighboring Mn layers are less favorable by 26.9-89.0 meV per formula unit (Table S1). Magnetic moment calculations provide strong supporting evidence, with Case 3's predicted value of 1.88 $\mu_B$ per atom showing excellent agreement with experimental measurements [1], compared to 2.31 $\mu_B$ and 2.19 $\mu_B$ for Cases 1 and 2 respectively. Experimental validation comes from multiple techniques: EDS measurements confirm the stoichiometric Cr : Mn ratio of 1 : 1.961 (Figs. S3a,b); XRD analysis demonstrates exceptional crystallinity with a sharp (004) peak (FWHM = 0.034°, Figs. S3d, e) and phase purity (Fig. S4); while SEM and atomic-resolution HAADF-STEM with EDS mapping (Figs. S5-S7) directly visualize the preferential Cr occupation of $Mn_1$ sites. This multi-technique approach establishes an unambiguous structure-property relationship, showing how single-layer Cr substitution governs both the structural and magnetic characteristics of this intriguing kagome material.

**Table S1**: The total energies of possible atoms arrangements in Fig. S2, with that of the ground state being set as zero.

| Possible configurations | Total energies per formula unit |
|---|---|
| Structure (a) | 0 meV |
| Structure (b) | 52.7 meV |
| Structure (c) | 89.0 meV |
| Structure (d) | 88.7 meV |
| Structure (e) | 26.9 meV |
| Structure (f) | 83.7 meV |

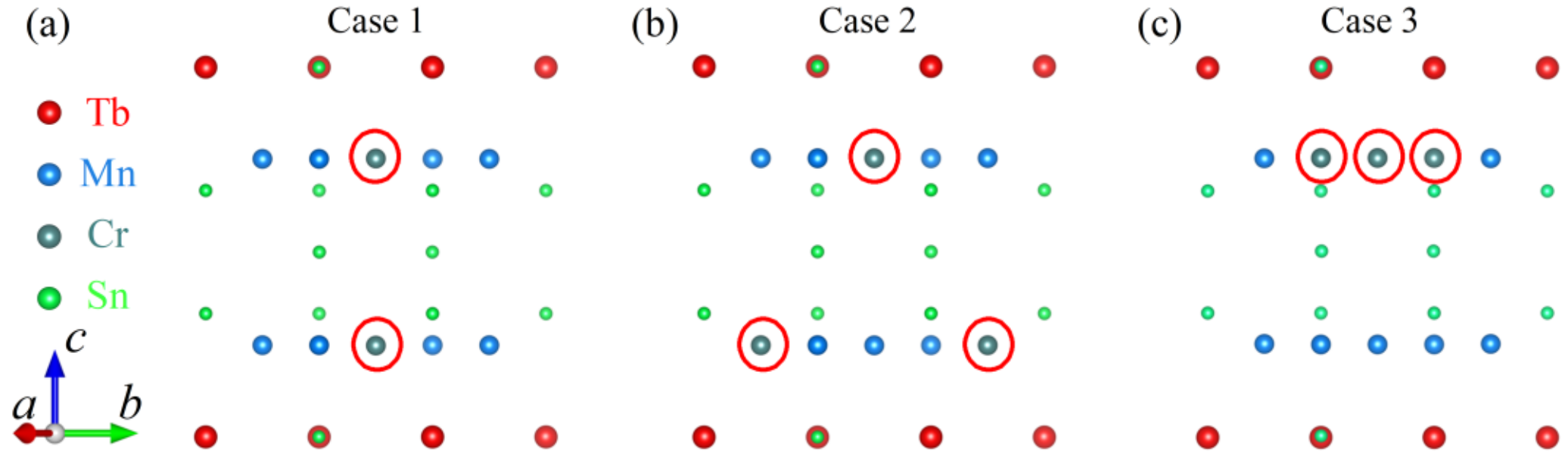

**Figure S1.** (a)-(c) Three possible cases for Cr atoms doping.

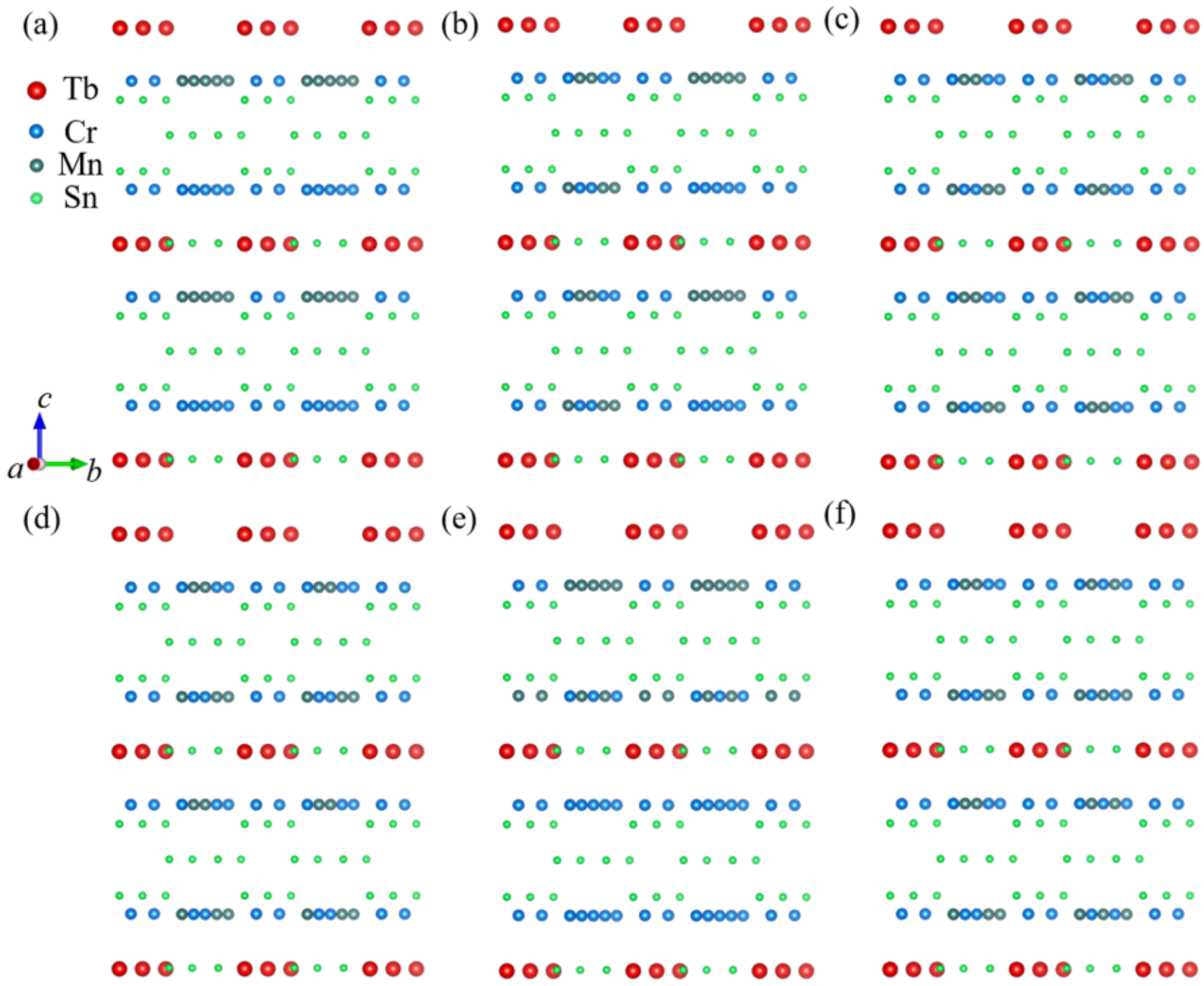

**Figure S2.** (a)-(f) The possible atoms arrangements in bulk $TbCr_2Mn_4Sn_6$.

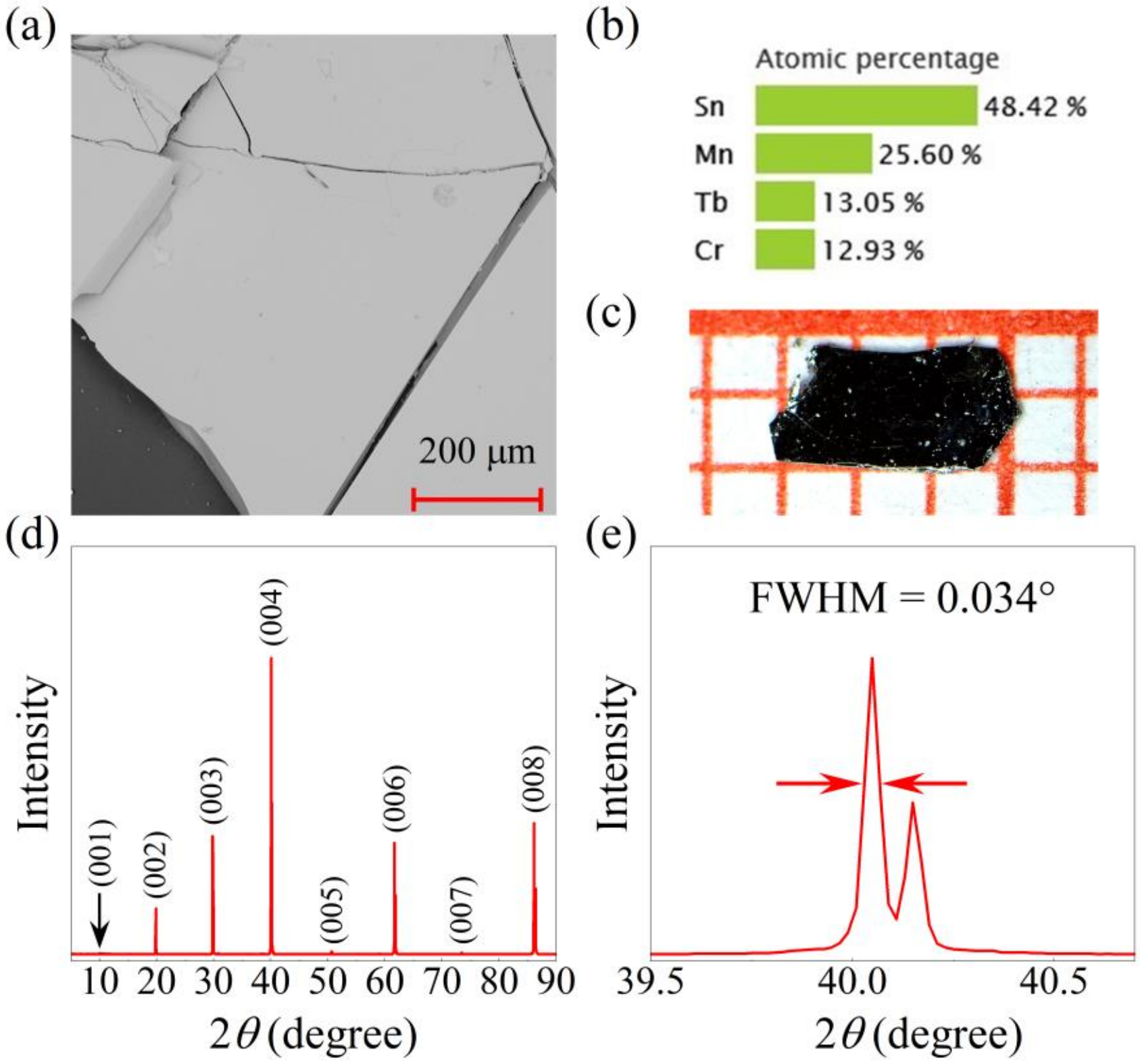

**Figure S3.** (a) and (b) The compositions of $TbCr_2Mn_4Sn_6$ crystal measured by EDS. (c) Picture of a typical $TbCr_2Mn_4Sn_6$ crystal. (d) The (00l) Bragg peaks in the Powder X-ray diffraction for $TbCr_2Mn_4Sn_6$ single crystals. (e) Powder X-ray diffraction patterns of the (004) Bragg peak for $TbCr_2Mn_4Sn_6$ single crystals. The full width at half maximum (FWHM) is 0.034°.

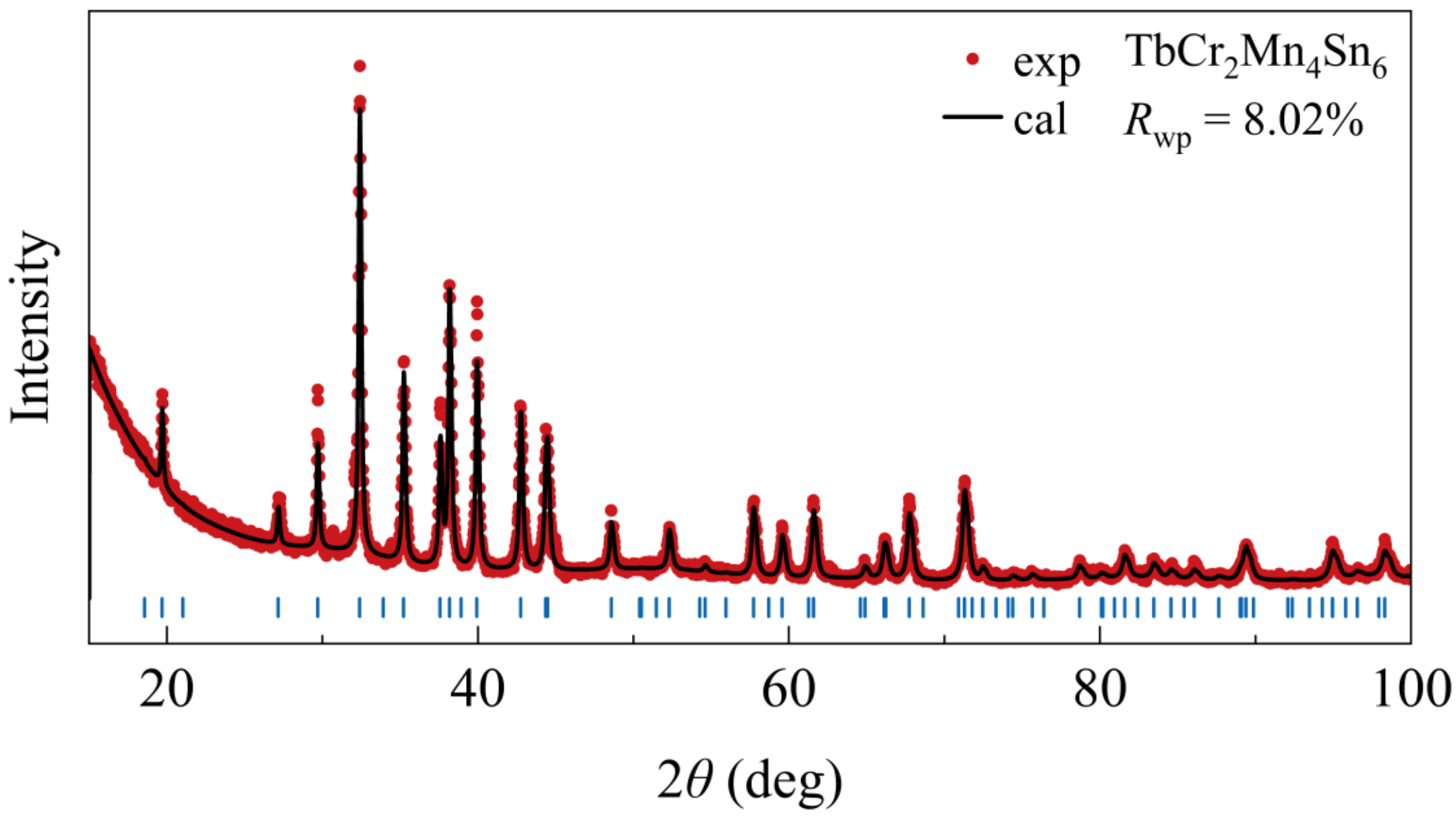

**Figure S4.** Rietveld refinement results of powder XRD of the $TbCr_2Mn_4Cr_2Sn_6$.

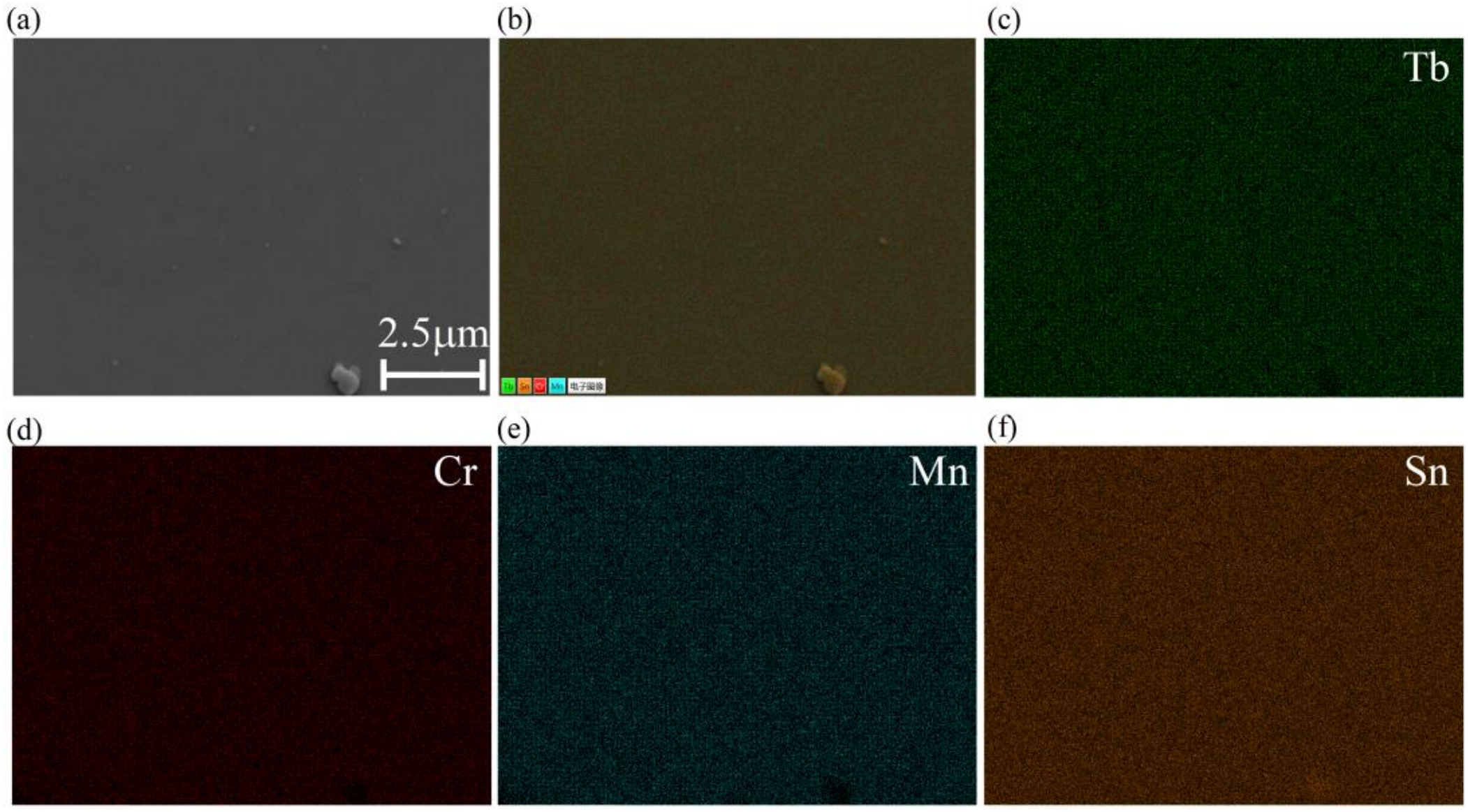

**Figure S5.** (a) SEM secondary electron image of the (001) plane in $TbCr_2Mn_4Sn_6$. Element mapping analysis in the scanning area: (b) all detected elements, (c) Tb, (d) Cr, (e) Mn and (f) Sn.

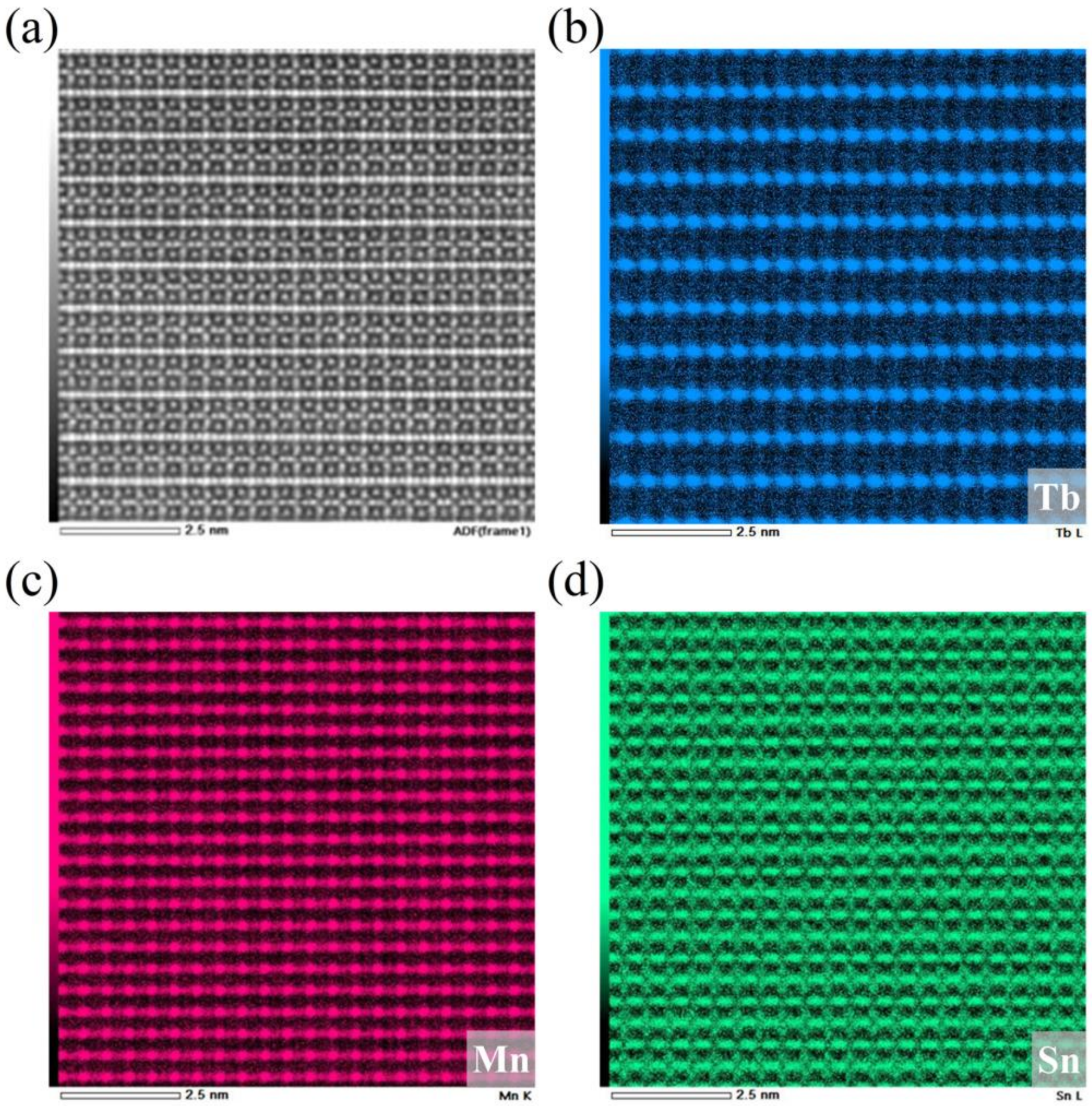

**Figure S6.** (a) The Cs-corrected HAADF-STEM original image of $TbCr_2Mn_4Sn_6$ crystals with EDS mapping results of Tb (b), Mn (e) and Sn (c) elements within the (110) plane. The scale bar in the image is 2.5 nm.

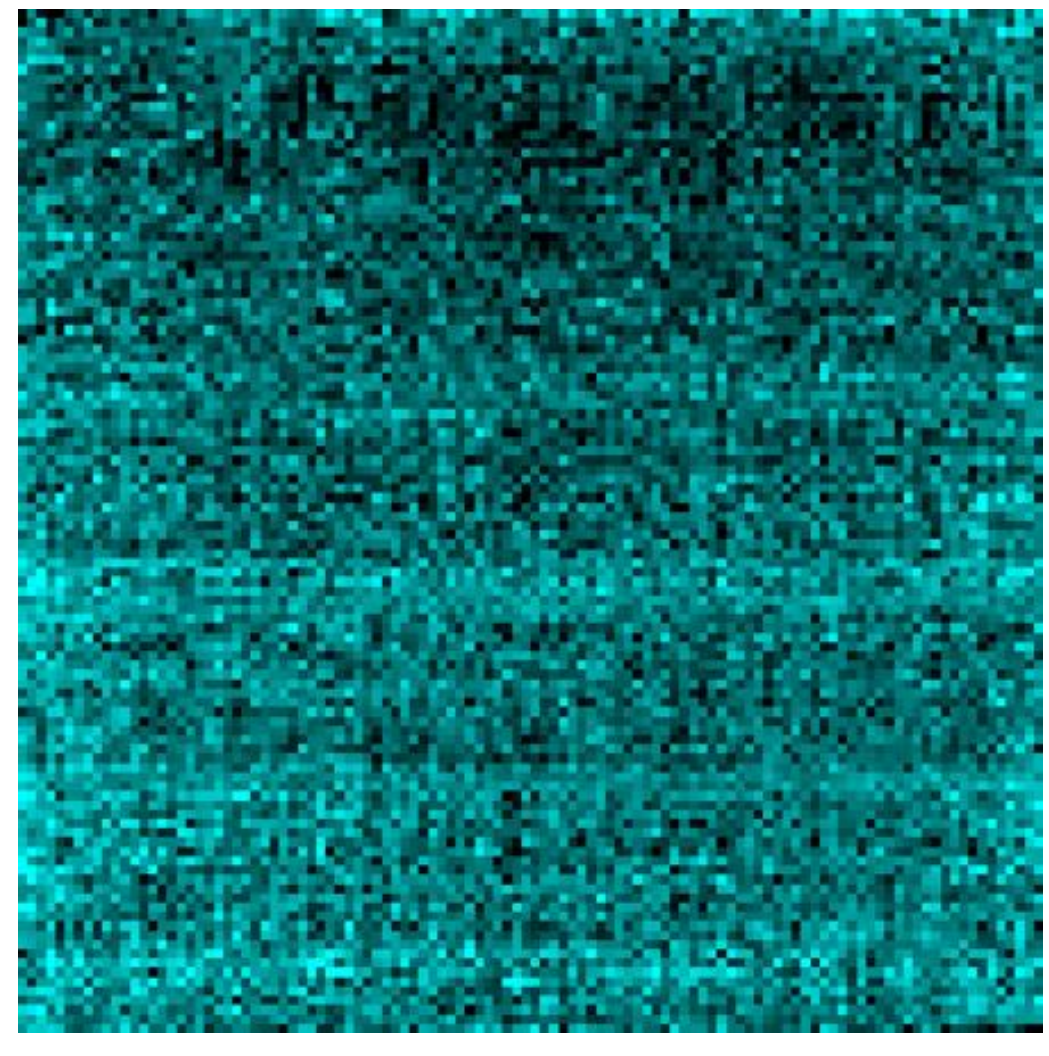

**Figure S7.** The Cr element mapping corresponds to the original HAADF image.

**Supplementary Note 2: Magnetic and magnetotransport measurements**

Figure S8 presents a systematic investigation of the temperature evolution (2-300 K) of Hall resistivity ($\rho_{xy}$) in Sample 1 under out-of-plane magnetic fields ($B$ // $c$-axis), revealing three distinct regimes: (1) Below 150 K, we observe antiferromagnetic (AFM) coupling between Mn and Tb layers along the $c$-axis with developing net magnetization and temperature-dependent hysteresis loops; (2) Between 150-220 K, the hysteresis gradually disappears while multiple low-field kinks emerge (Fig. S8b), indicating competing magnetic phases; (3) Above 220 K (Fig. S8c), only the anomalous Hall signal persists as the field aligns in-plane moments along the $c$-axis. Complementary magnetization measurements on Sample 2 (Fig. S9, 160-300 K) and detailed transport studies on Sample 3 (complete data in Fig. S10, selected data in main text Fig. 3a) consistently confirm these phase transitions. Component-resolved analysis (Fig. S11) quantitatively distinguishes the total ($\rho_{yx}$), normal ($\rho^{N}_{yx}$), anomalous ($\rho^{A}_{yx}$), and topological ($\rho^{T}_{yx}$) Hall contributions, with magnified views (Figs. S11b,d) demonstrating reproducible transition points across measurements. Additional characterizations include temperature-dependent longitudinal resistivity (Fig. S12a), extracted anomalous Hall resistivity (Fig. S12b), and comparative analysis of maximal topological Hall resistivity with literature values (Table S2 [2-35]). This comprehensive dataset not only validates the main text findings but also provides deeper insights into the temperature evolution of magnetotransport properties, establishing clear phase boundaries and quantitatively benchmarking the observed topological Hall effects against known systems.

**TABLE S2**: The maximal topological Hall resistivity observed in various compounds.

| Compounds | Maximal $\rho^{T}_{xy}$ ($\mu\Omega \cdot$ cm) | Refs |
|---|---|---|
| MnSi | 0.0045 | [2] |
| MnBi | 0.021 | [3] |
| $Mn_5Si_3$ | 0.05 | [4] |
| $Cr_{1.53}Te_2$ | 0.106 | [5] |
| $EuAl_4$ | 0.13 | [6] |
| MnNiGa | 0.15 | [7] |
| MnGe | 0.16 | [8] |
| $(Cr_{0.9}B_{0.1})Te$ | 0.21 | [9] |
| $SmMn_2Ge_2$ | 0.3 | [10] |
| $CeMn_2Ge_2$ | 0.8 | [11] |
| $Fe_{0.7}Co_{0.3}Si$ | 0.82 | [12] |
| $Mn_{1.4}PtSn$ | 0.9 | [13] |
| $Cr_{2.76}Te_4$ | 1.1 | [14] |
| $Fe_{0.28}TaS_2$ | 1.4 | [15] |
| $Mn_2PtSn$ | 1.53 | [16] |
| SmAlSi | 1.7 | [17] |
| $YMn_6Sn_6$ | 2 | [18] |
| $Mn_2CoAl$ | 2 | [19] |
| $Mn_{2-x}Zn_xSb$ | 2 | [20] |
| $Fe_3Sn_2$ | 2.01 | [21] |
| $Fe_3GeTe_2$ | 2.04 | [22] |
| $NdMn_2Ge_2$ | 2.05 | [23] |
| $ErMn_6Sn_6$ | 2.09 | [24] |
| $Fe_5Sn_3$ | 2.12 | [25] |
| $Cr_{1/3}NbS_2$ | 2.33 | [26] |
| $Gd_2PtSi_3$ | 2.61 | [27] |
| $Mn_3Sn$ | 3.13 | [28] |
| $LaMn_2Ge_2$ | 4.5 | [29] |
| $Fe_3Ga_4$ | 4.83 | [30] |
| $Mn_2FeSn$ | 5.19 | [31] |
| $MnBi_4Te_7$ | 7 | [32] |

| EuAgAs | 7 | [33] |
|---|---|---|
| EuCuAs | 7.4 | [34] |
| EuO | 12 | [35] |
| $TbCr_2Mn_4Sn_6$ | 19.1 | This work |

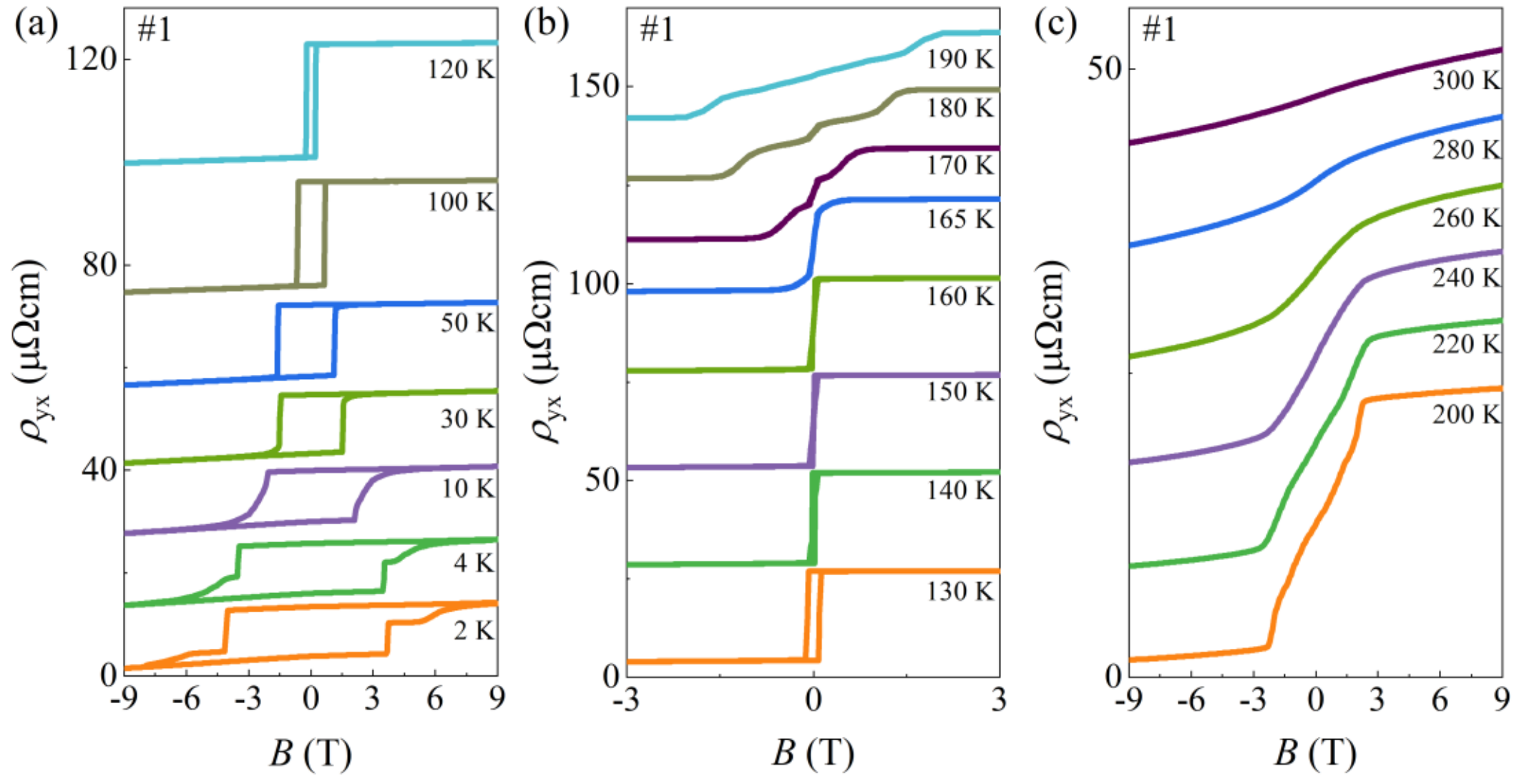

**Figure S8.** (a)-(c) Hall resistivity ($B$ // $c$-axis) as a function of magnetic field $B$ at 2-300 K for sample 1.

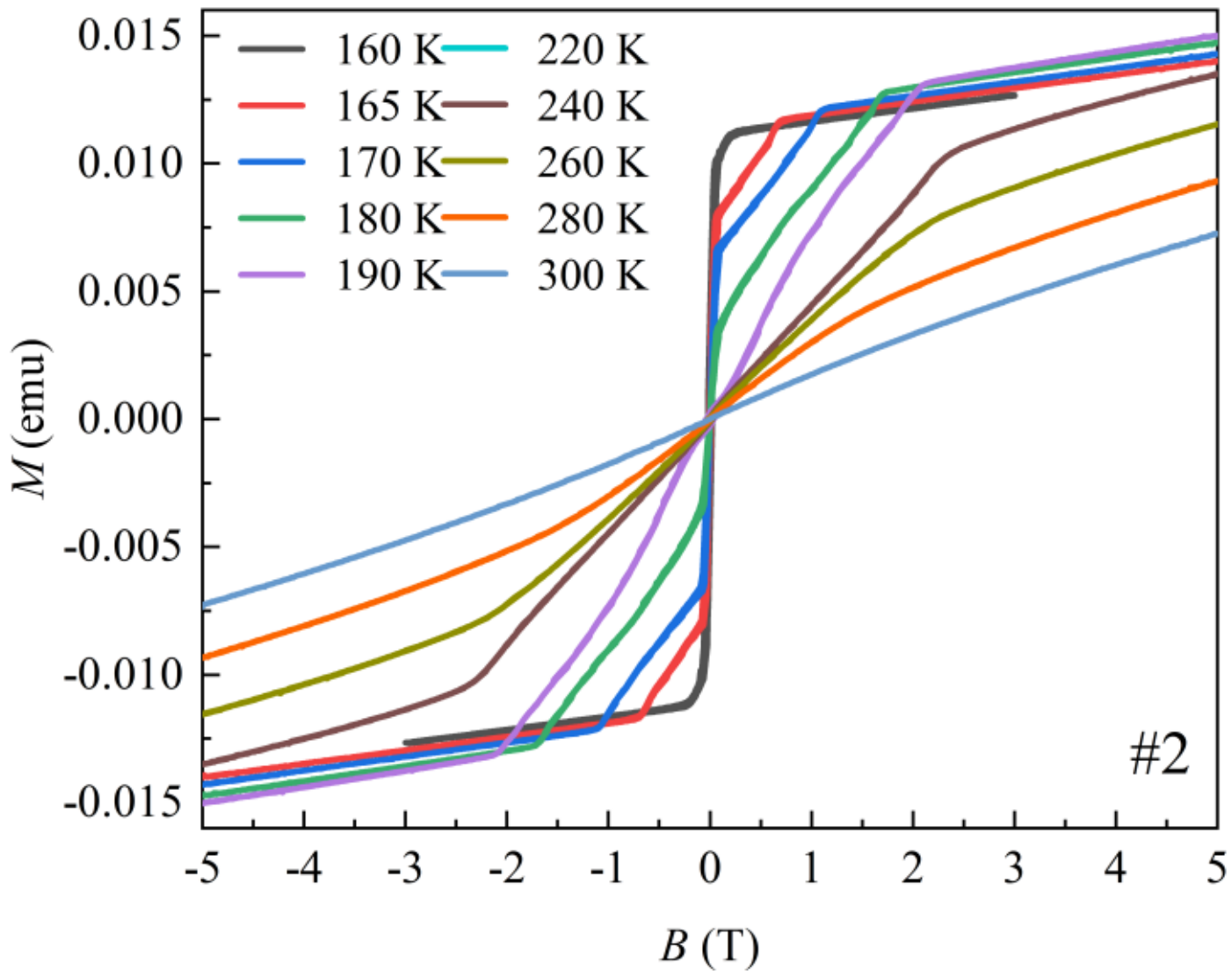

**Figure S9.** Isothermal magnetization ($B$ // $c$-axis) as a function of $B$ at different temperatures for Sample 2.

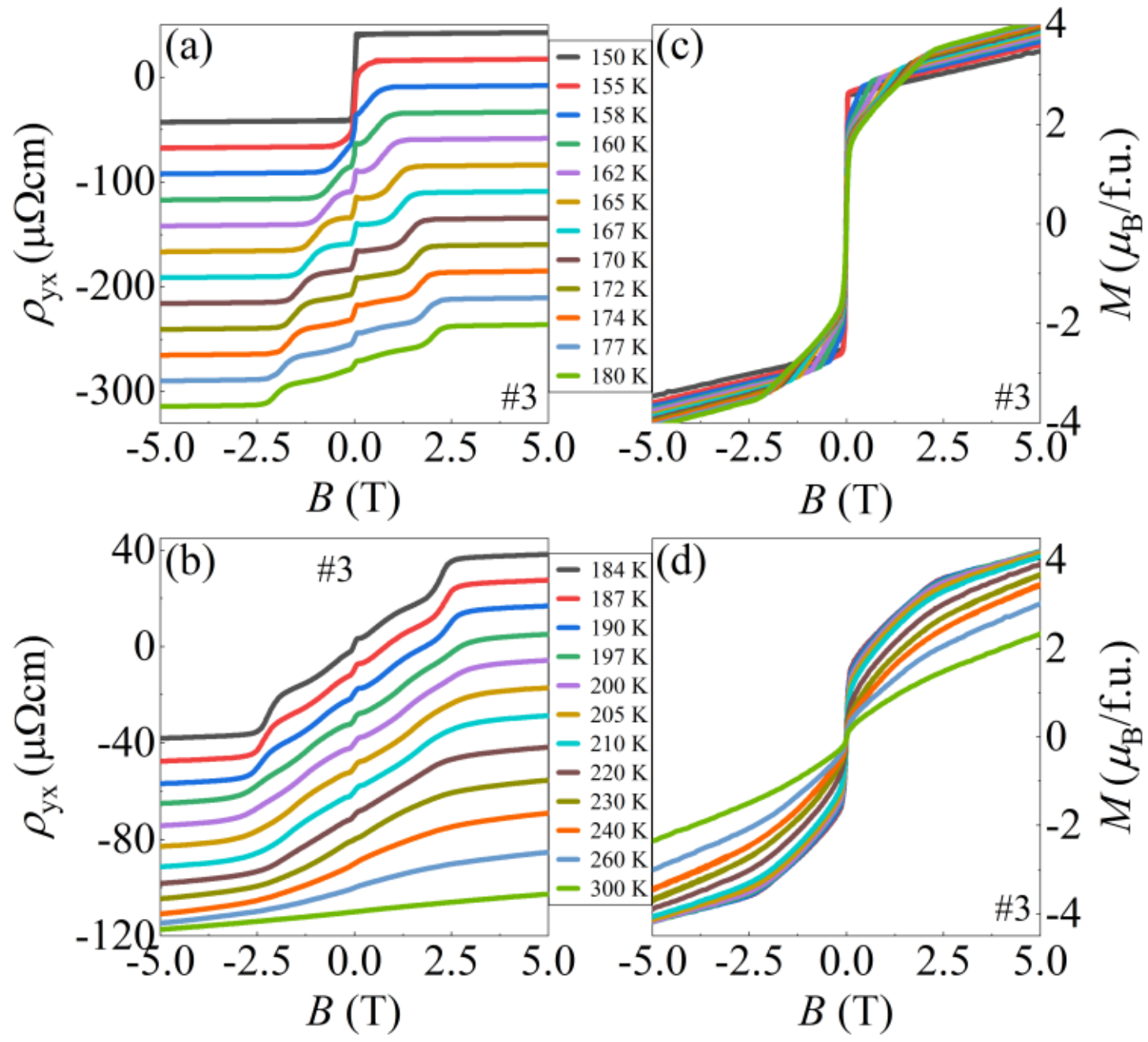

**Figure S10.** (a)-(b) Hall resistivity as a function of $B$ at different values of temperature with the magnetic field along the $c$-axis for Sample 3. (c)-(d) Isothermal magnetization as a function of $B$ at different values of temperature with the magnetic field along the $c$-axis for Sample 3.

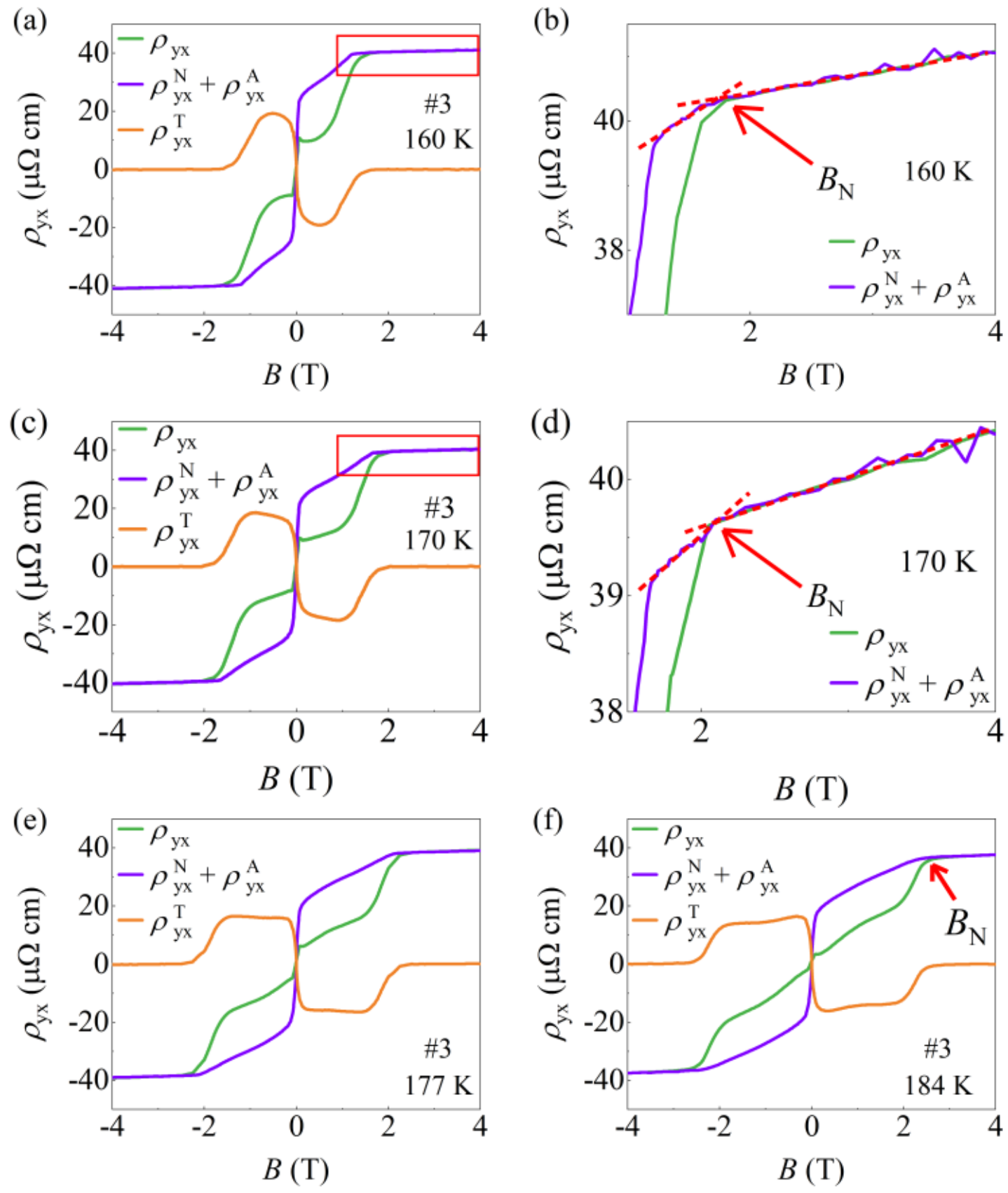

**Figure S11.** The representative $\rho_{yx}$ curve (green), calculated $R_0B + 4\pi R_S M$ fitting curve (purple) and extracted $\rho^T_{yx}$ curve (orange) for Sample 3 measured at (a) 160 K, (c) 170 K, (e) 177 K, and (f) 184 K, respectively. (b) and (d) are magnified views of the red-boxed regions in panels (a) and (c), respectively.

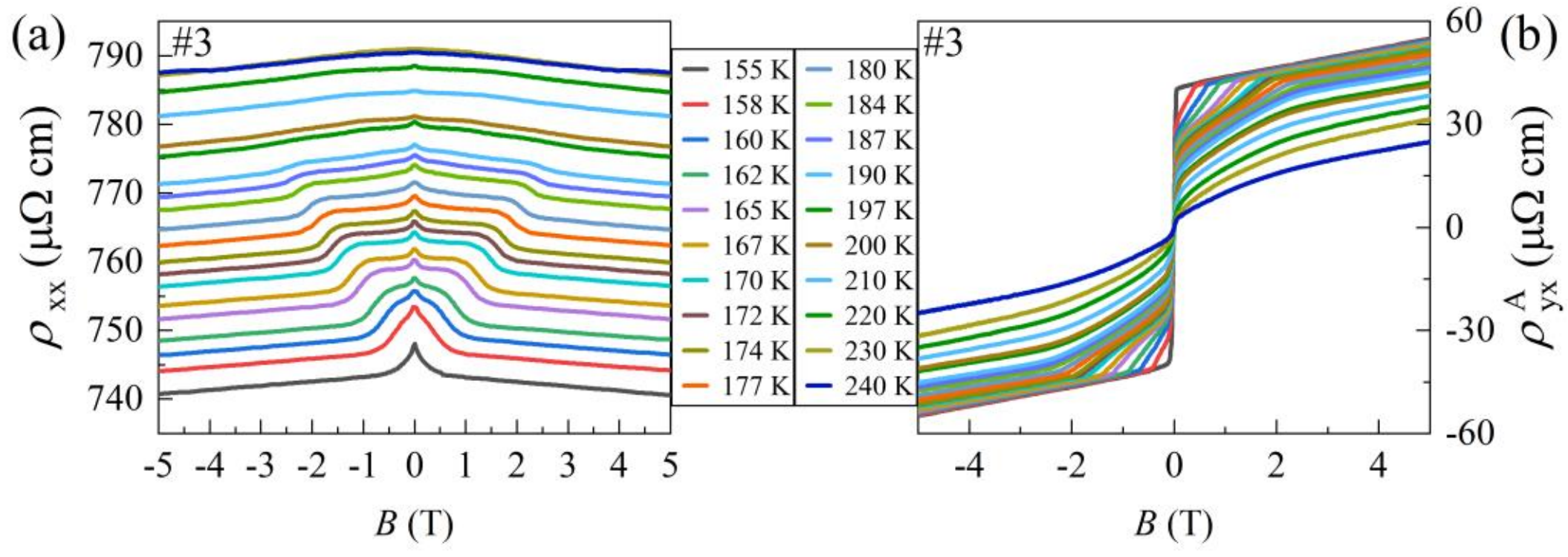

**Figure S12.** $\rho_{xx}$ as a function of $B$ at various temperatures with the magnetic field along the $c$-axis for crystal 3. (b) Anomalous Hall resistivity $\rho^A_{yx}$ ($B$ // $c$-axis) as a function of $B$ at different temperatures for crystal 3.

**Supplementary Note 3: Magnetic force microscopy measurements**

Our magnetic force microscopy (MFM) investigations (Figs. S13-S16) provide comprehensive characterization of domain behavior across field and temperature variations. Beginning with a pristine 70 $\mu$m × 70 $\mu$m FIB-milled surface (Fig. S13), we developed a robust methodology for quantifying domain widths and intensities (Fig. S14), enabling systematic analysis presented in the main text Figs. 4(p,q). Remarkably, field-dependent studies at 180 K demonstrate perfect magnetic reversibility, with positive (2.45 T→0.01 T, Fig. S15a) and negative (-0.05 T→-2.5 T, Fig. S15b) field sequences showing identical domain configurations at equivalent field strengths, quantitatively confirmed through width analyses (Figs. S15c,d). This polarity independence reveals fundamentally symmetric domain energetics without hysteresis. Temperature-dependent measurements (150-220 K, Fig. S16) track domain evolution, with purple circles and red squares respectively marking critical fields for width transitions and domain disappearance (scale bar: 2 $\mu$m). Together, these studies establish three key findings: (1) complete field-polarity symmetry in domain configurations, (2) temperature-dependent shifting of transition fields without morphological changes, and (3) well-defined critical points throughout the studied range, providing a complete picture of the system's magnetic domain physics.

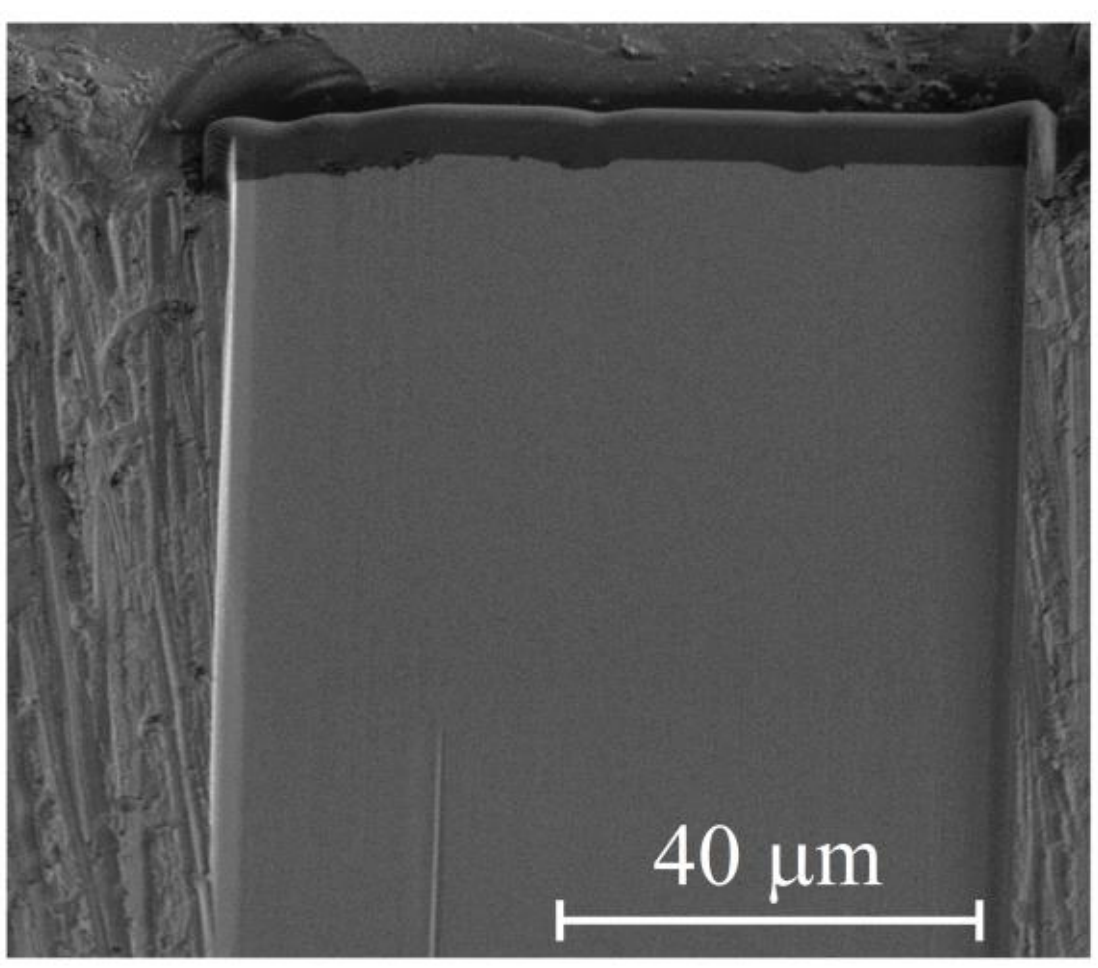

**Figure S13.** MFM measurement on a focused ion beam (FIB) prepared 70 μm × 70 μm surface.

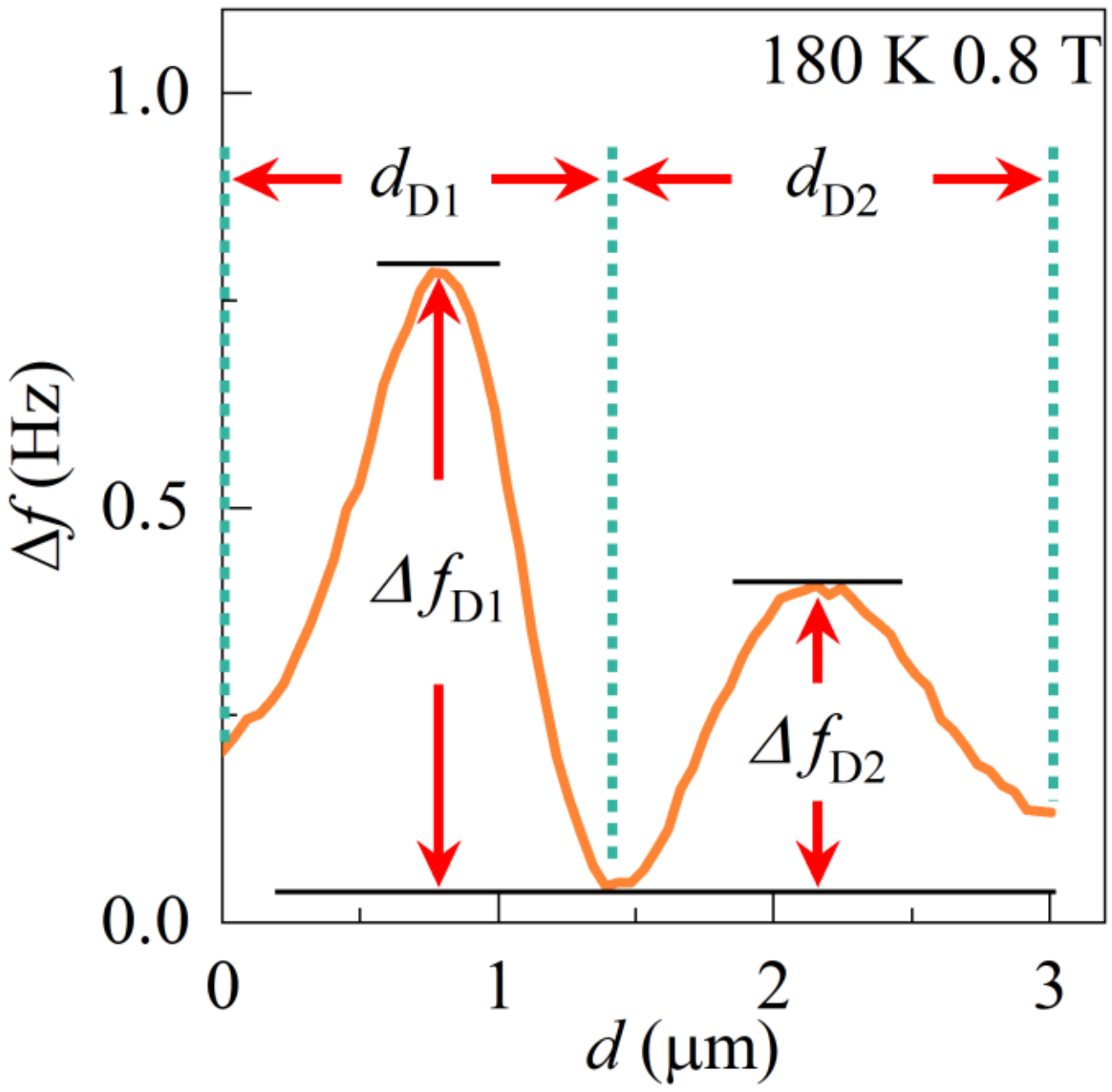

**Figure S14.** MFM signals measured along the blue lines marked in Fig. 4(d) under 0.8 T at 180 K. $\Delta f_{D1}$ and $\Delta f_{D2}$ denote the signal intensities of magnetic domains $D_1$ and $D_2$, respectively, while $d_{D1}$ and $d_{D2}$ represent their corresponding widths.

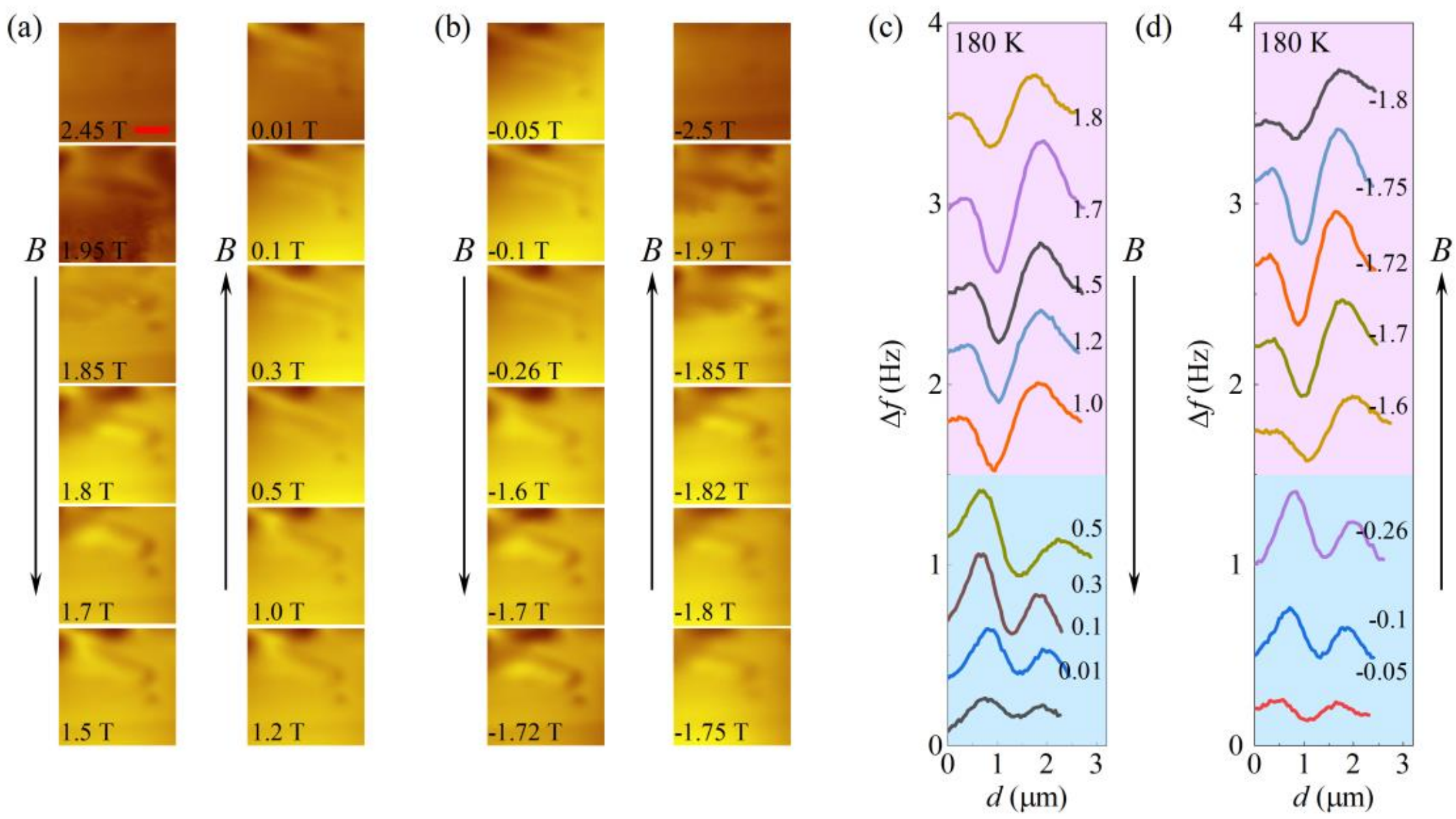

**Figure S15.** MFM images were taken at 180 K on the same location of $TbCr_2Mn_4Sn_6$ with (a) decreasing magnetic fields of 2.45 to 0.01 T, (b) increasing magnetic fields of -0.05 to -2.5 T. The scale bars attached to (a) and (b) are both 2 μm. (c) and (d) MFM signals measured along the blue lines marked in Fig. 4(d) under different magnetic fields at 180 K, corresponding to panels (a) and (b) respectively.

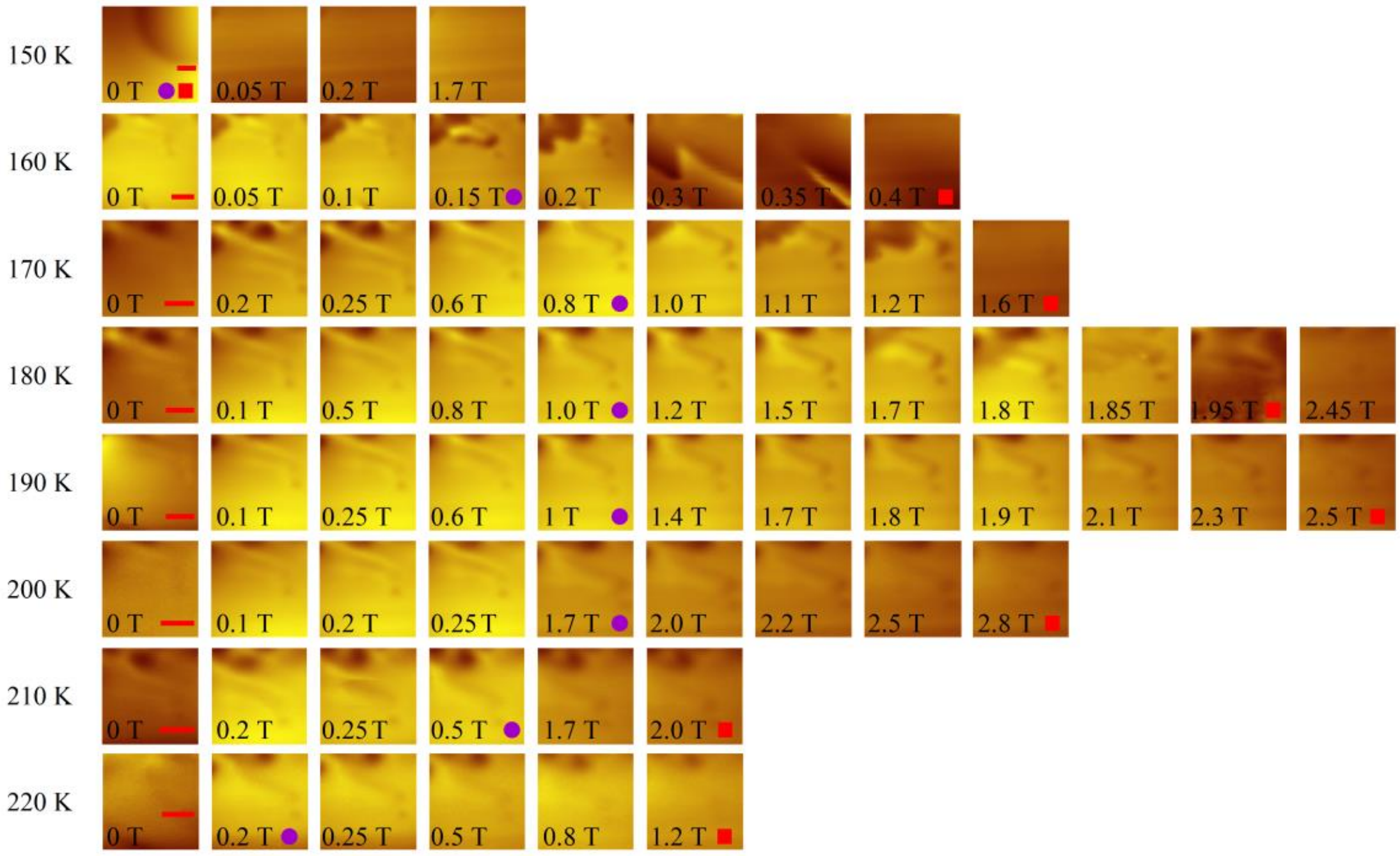

**Figure S16.** MFM images measured under various magnetic fields from 150 K to 220 K, showing critical magnetic fields for magnetic domain width transitions (purple

circles) and magnetic domain disappearance (red squares).

## Supplementary Note 4: First-principles calculation

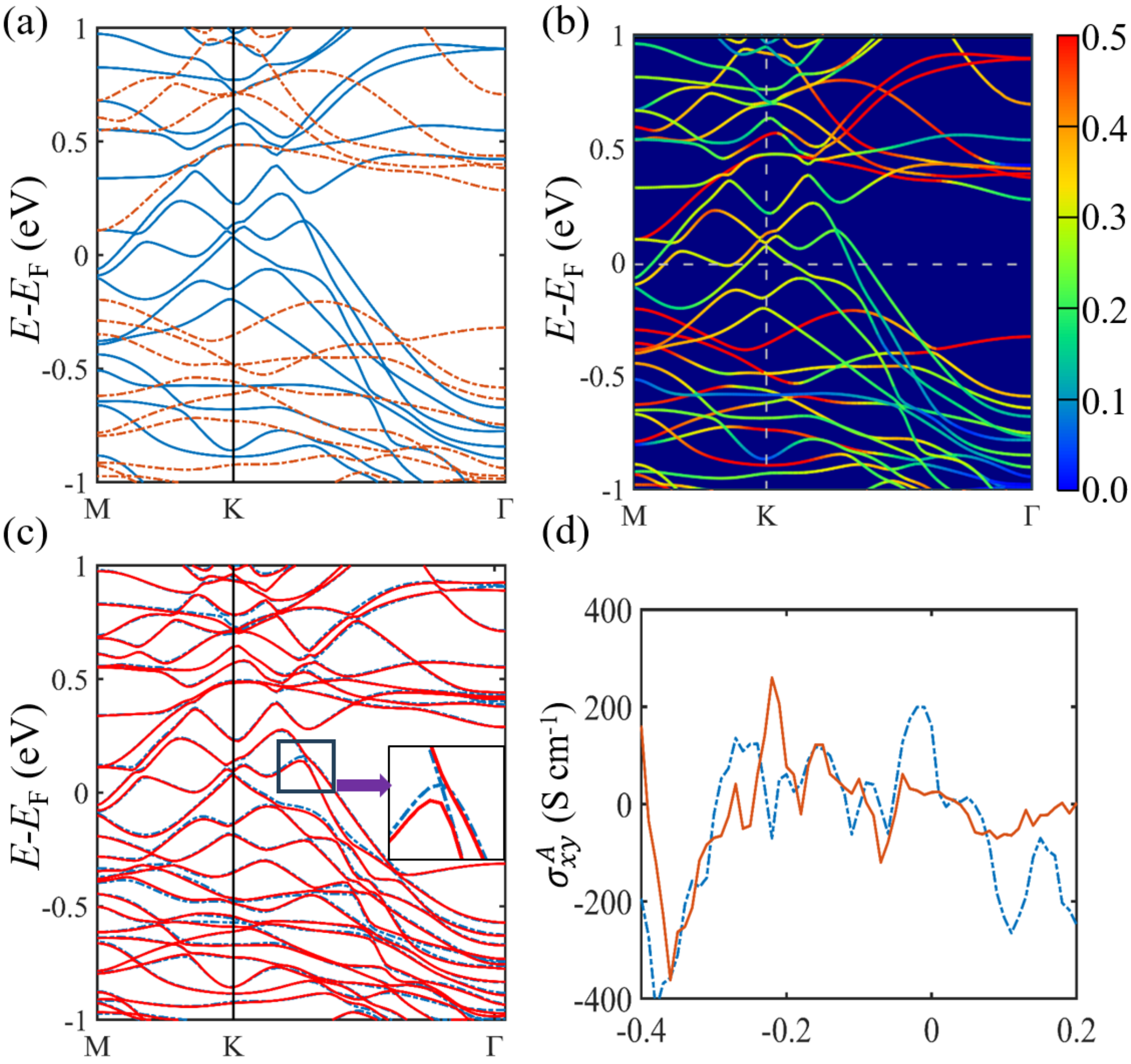

**Figure S17.** (a) The electronic structure of $TbCr_2Mn_4Sn_6$. (Blue: spin up, Red: spin down). (b) Contribution from Cr atom with SOC (Mn spin along out-of-plane direction). (c) The energy bands near the Fermi level of $TbCr_2Mn_4Sn_6$ can tuned by Mn spin reorientation, as marked by the black box. Blue: Mn spin along out-of-plane direction. Red: Mn spin along in-plane direction. (d) The calculated anomalous Hall conductivity $\sigma^{A}_{xy}$ of $TbCr_2Mn_4Sn_6$.

Fig. S17 presents a detailed investigation of the electronic structure modifications and anomalous Hall effect (AHE) in $TbCr_2Mn_4Sn_6$. Panel (a) displays the spin-resolved band structure, where blue and red lines correspond to spin-up and spin-down channels respectively, revealing significant spin polarization. The introduction of Cr atoms, as shown in panel (b) with spin-orbit coupling (SOC) included for out-of-plane Mn spin alignment, demonstrates substantial orbital hybridization while simultaneously breaking the kagome lattice symmetry and reducing the point group symmetry from $D_{6h}$ to $C_2$. Panel (c) highlights the remarkable sensitivity of Fermi level electronic states to Mn spin orientation, comparing out-of-plane (blue) and in-plane (red) alignments, with the black circle marking bands undergoing significant reconstruction during spin reorientation. Most importantly, panel (d) establishes that the anomalous Hall conductivity ($\sigma^A_{xy}$) originates primarily from accidental band crossings induced by the combined effects of symmetry breaking and strong spin-orbit interactions. These findings collectively demonstrate how Cr doping transforms both the structural symmetry and electronic landscape, creating conditions for (1) spin-texture-dependent band reconstruction, (2) symmetry-protected accidental crossings, and (3) consequent large intrinsic AHE - all stemming from the reduced $C_2$ symmetry and its influence on Berry curvature distribution in the system.